\documentclass[journal]{IEEEtran}
\usepackage{graphicx}
\usepackage{epstopdf} 
\usepackage{times,epsfig,latexsym,amssymb,psfrag}
\usepackage{paralist,citesort}
\usepackage{tikz,pgf,pgfplots}
\usepackage{tikz}
\usepackage{lipsum,adjustbox}
\usetikzlibrary{mindmap,calc,positioning}
\usepackage{booktabs}
\usepackage{subcaption}
\usepackage{caption}
\usepackage{eurosym}
\usepackage{float}
\usepackage{mathrsfs}
\usepackage{enumitem}
\usepackage{amsmath}
\usepackage{mathtools}
\usepackage{xcolor}
\usepackage[edges]{forest}
\usepackage[ruled,vlined]{algorithm2e}
\IEEEoverridecommandlockouts
\usepackage{soul}

\title{Ultra-Low-Power IoT Communications: A novel address decoding approach for wake-up receivers}
\author{ 
{Yousef Mafi},
 {Fakhreddin Amirhosseini}, 
 {Seied Ali Hosseini}, 
 {Amin Azari}, 
 {Meysam Masoudi,
 {Mojtaba Vaezi}}
 \\
{KTH Royal Institute of Technology, Sweden (aazari@kth.se)} \\}
\IEEEpeerreviewmaketitle

\begin{document}
	\setlength{\textfloatsep}{5pt}
    \SetKwComment{Comment}{}{}
	\maketitle
	\thispagestyle{empty}
	\pagestyle{empty}
	\begin{abstract}
	 Providing energy-efficient Internet of Things (IoT) connectivity has attracted significant attention in fifth-generation (5G) wireless networks  and beyond.  A potential solution for realizing a long-lasting network of IoT devices is to equip each IoT device with a wake-up receiver (WuR) to have always-accessible devices instead of always-on devices.  WuRs typically comprise a radio frequency demodulator, sequence decoder, and digital address decoder and are provided with a unique authentication address in the network. Although the literature on efficient demodulators is mature, it lacks research on fast,
low-power, and reliable address decoders. As this module continuously monitors the received ambient energy for potential paging of the device, its contribution to WuR's power consumption is crucial. Motivated by this need, a low-power, reliable address decoder is developed in this paper. We further investigate the integration of WuR in low-power uplink/downlink communications and,  using system-level energy analysis; we characterize operation regions in which WuR can contribute significantly to energy saving.   The device-level energy analysis confirms the superior performance of our decoder. {The  results show that the proposed decoder significantly outperforms the state-of-the-art with power consumption of $60$ nW, at cost of compromising negligible increase in decoding delay.}

	\end{abstract}
	

\section{Introduction}
The Internet of Things (IoT) use cases, such as massive machine-type communications (mMTC) and ultra-reliable low-latency communications (URLLC), are one of the key drivers of fifth-generation (5G) wireless networks and beyond (6G) \cite{ref101,7giot}.
A full realization of IoT can make substantial changes the life quality by providing an intelligent network covering every object in the daily life, such as in agriculture, surveillance, health care, body implants, home automation, etc \cite{ref1,ref2}.  Each IoT node is typically equipped with sensors, microcontrollers, wireless transceivers, and energy sources such as batteries. The principal objective of most reporting IoT devices is to observe their surroundings and report/act upon their gatherings. Since most IoT devices are battery-powered, and once deployed, they are expected to live a long time without human intervention, battery lifetime has always been a key concern of IoT-enabled networks \cite{5ga}. Hence,  the IoT traffic demands support for low-energy consumption.
\vspace{-11pt}
\subsection{Energy-Efficient  Cellular IoT}
During the evolution of mobile networks towards 5G and beyond, the energy efficiency of large-scale IoT communications have been improved significantly \cite{5ga,nbt,nbtA}. The implemented methods and solutions to realize the energy-efficient IoT communications are summarized in \cite{azari2018serving,9017997}. As per literature, these solutions can be categorized into evolutionary and revolutionary solutions~\cite{revol}. The evolution began from LTE Release 12 by supporting low-cost communications over dedicated resources followed by in Release 13 \cite{nok1} where  LTE-MTC (LTE-M) and narrow-band IoT (NB-IoT) systems were introduced. NB-IoT is known as a revolutionary solution and reference signals and random access happen in time instead of frequency \cite{mnbt,nbtA}. Since in these systems the link budget is improved, the required energy to transmit a bit is reduced~\cite{masoudi2021low,z5giot,znbt}. However, the access procedures requires listening to the control channel and exchanging signal for random access, synchronization, and resource reservation~\cite{znbt,nbt}, resulting in a challenge to realize the long-lasting battery limited devices activities ~\cite{z5giot,myp}. As per the state-of-the-art, extensive efforts are made to improve the energy efficiency of  IoT devices such as discontinuous reception (DRX) \cite{maheshwari2018drx}, power-saving mode (PSM) \cite{gsm2019nb}, lightweight communications protocols \cite{ref3,ref4},  low-power radio transceivers \cite{ref3,ref5}, idle listening \cite{refa5}, and duty cycling \cite{ref4,refb1}. These methods imply an inherent trade-off between reachability and energy saving of the devices \cite{ze}. Motivated by addressing this challenge, the energy/reachability trade-off can be broken by leveraging device hibernation, and wake-up receivers (WuR) \cite{ze}. 
\vspace{-10pt}
\subsection{The WuR Approach}
In WuR approach, the device is in hibernation mode unless it is paged by the base station (BS)   or has a message to exchange. Since the battery is entirely intercepted as it is hibernating \cite{refa2}, no power is consumed except for monitoring whether the device is paged. To page a device, the network broadcasts a wake-up signal, including the device's unique address, to all devices. The device is equipped with an auxiliary WuR--consuming much less power than the primary device \cite{refa4}--to receive and decode every wake-up signal  and prompt the device if required \cite{refa5,refa6}.
As a result, the power consumption of WuR lies in the $\mu$W domain, whereas the IoT device may consume several milliwatts \cite{refa7,refa8}. Therefore, reliable self-sufficient energy management is indispensable for achieving an eligible WuR capable of operating indefinitely without compromising the quality of service (QoS) \cite{ref3}. A WuR contains a radio frequency (RF) receiver, a demodulator, a low noise baseband amplifier, and a digital address decoding unit. In this regard, ultra-low-power,  noise-robust address decoders are presented to enhance WuR's power efficiency. {For such ultra-low-power devices, harvesting energy} {from the  environment can provide enough power, e.g., around 1$\mu$W, to  keep WuRs charged  to operate self-sustainably}\footnote{{The  energy can be harvested from 1) environment such as solar and wind, 2) the ambient wireless radio waves and signals such as TV broadcasts 3)  other sources like temperature difference, and motion. The interested readers are referred to} \cite{sudevalayam2010energy,sanislav2021energy,shin2017cooperative} {for in-detail description of the harvesting technologies, the conversion power efficiency, and the sources of energy harvesting. }} \cite{ref8,ref9,ref11}.
\vspace{-10pt}
\subsection{Related Works}
\subsubsection{WuR integration in cellular networks}
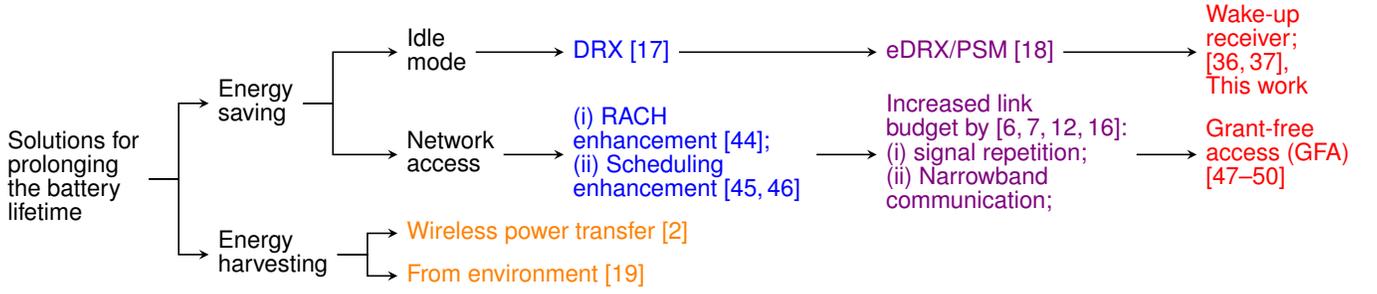
\begin{figure*}
 	\begin{forest}
 		for tree={
 			align=left,
 			edge = {draw, semithick, -stealth},
 			anchor = west,
 			font = \small\sffamily\linespread{.84}\selectfont,
 			forked edge,          
 			grow = east,
 			s sep = 0mm,    
 			l sep = 8mm,    
 			fork sep = 4mm,    
 			tier/.option=level
 		}
 		[
 		    Solutions for  \\ prolonging \\ the battery \\lifetime						
   			 	 [ Energy \\ harvesting [\textcolor{orange}{From environment \cite{ref3}} ] [\textcolor{orange}{Wireless power transfer \cite{7giot}} ]
 		 	 ]
 		 [Energy \\saving
 		                 [Network\\ access
                    [\textcolor{blue}{(i) RACH}\\ \textcolor{blue}{enhancement \cite{pang2015evaluation};} \\\textcolor{blue}{(ii) Scheduling}\\\textcolor{blue}{enhancement \cite{azari2015lifetime,azari2017network} }
                    [ \textcolor{violet}{Increased link }\\ \textcolor{violet}{budget by \cite{nbt,mnbt,nbtA,myp}:}\\\textcolor{violet}{(i) signal repetition;}\\\textcolor{violet}{(ii) Narrowband} \\\textcolor{violet}{communication;}\\
                    [\textcolor{red}{Grant-free} \\\textcolor{red}{access (GFA)} \\\textcolor{red}{\cite{gf,gf3,dln,m2018g}}]]]
                ]
                         		[Idle\\ mode
         		[\textcolor{blue}{DRX \cite{maheshwari2018drx}} [\textcolor{violet}{eDRX/PSM \cite{gsm2019nb}} [\textcolor{red}{Wake-up} \\\textcolor{red}{receiver;}\\\textcolor{red}{\cite{3gpp,3gpp2},}\\\textcolor{red}{This work}]]]
         		]
 		]
 		]
 	\end{forest}
 \caption{Classification of various solutions targeting  a longer battery lifetime: 4G  enhancements for IoT are in blue, LTE-Pro and 5G enhancements are in violet, proposed approaches for beyond  5G are in red, and the  device-dependent approaches are in orange. } \label{fig:class}
 \end{figure*}

The 3rd Generation Partnership Project (3GPP)  Release 15 introduces the wake-up signal as a unique signal sent from the BS to IoT devices paging them to wake the main receiver up \cite{3gpp,3gpp2}. In order to avoid creating extra interference in  orthogonal frequency-division multiplexing (OFDM)  across cellular networks, wake-up devices should use out-of-band (or in guard band) signal, as in the NB-IoT \cite{guard}. The applicability of leveraging WuR in millimeter wave (mmWave) bands and beamforming systems has been investigated in \cite{3gpp3}. In \cite{3gpp5}, the authors propose  a novel approach to transmit a wake-up signals along with the normal transmissions. In \cite{refa2}, a prototype implementation of a WuR-based two-tier system, bridging cellular IoT and Bluetooth low energy (BLE), has been presented. In this implementation, BLE devices are activated using BLE signals, and an 8-bit address comparator module \cite{3gpp9} is leveraged in the WuR for address decoding and paging verification. In \cite{3gpp7}, the performance of  WuR-enabled cellular IoT in uplink and downlink  has been investigated. The authors derived  lower and upper bounds for the success and false wake-up probabilities as a function of signal-to-interference-plus-noise ratio (SINR).  In \cite{ze}, a joint WuR and energy-harvesting connectivity solution have been proposed in which information packets sent on the proposed solution are preceded by a power-optimized preamble. Then, the energy harvested from the preamble at the WuR is used by a low-power receiver for decoding the data packets.  In Fig. \ref{fig:class}, the approaches aiming at enhancing the battery lifetime of devices are grouped. In this figure, 4G enhancements are in blue, LTE-advanced/pro and 5G enhancements in purple, and proposals for beyond  5G approaches in red. The device-dependent approaches are in orange, i.e., wireless power transfer from a charging drone to remote IoT devices \cite{uavCh}.  

\subsubsection{WuR device design}
Address decoders, alongside RF receivers and demodulators, consume a considerable portion of the overall power. Although the literature on making efficient demodulators is mature,  the literature lacks research on fast yet low-power and reliable address decoders. The majority of proposed WuRs in the literature leverage microcontrollers or digital correlators as their address decoders, as per \cite{refa2,ref14,ref15}, and these modules consume a considerable portion of the total power in lengthy addresses used in local and wide-area deployments. The authors in  \cite{ref13} have utilized an off-chip lumped element matching network for the wake-up receiver.  The study in  \cite{ref14} has utilized PIC12F683 MCU, a 3-volt prototype decoder that consumes 820uW during interception. AS393X commercial comparator chips have also been utilized for address decoding, which has a built-in on-off keying (OOK) demodulator, envelope detector, and address decoder unit was consuming 3.9$\mu$W during interception. The authors in \cite{ref17} have designed a switch-capacitor-based correlation unit inside an ASIC which draws 0.4$\mu$A at 20kS/s (kilo samples per second). The studies in \cite{ref16,ref18,ref19} have utilized flip flops (FF) based decoder to generate a wake-up interrupt signal which consumes 13.4$\mu$W at 0.9 volts and 100Kbps and decodes a 16-bit address in 0.08ms. Moreover, they have presented pipelined FFs to store intercepted signal bits for real-time address validation using logic gates.


\vspace{-11pt}
\subsection{Motivation and Contributions}
The increase in the number of connected devices and the need for security of IoT communications mandate adopting lengthier addresses, which increases the energy consumption of the address decoding module in the WuR.  In order to save energy in continuously monitoring the received ambient engines for potential paging,   a  low-power yet reliable address decoder is needed. Here, a novel address decoder architecture is proposed and evaluated in terms of power efficiency and reliability.
The major contributions of this study include:
\begin{itemize}
\item We propose a novel address decoder for WuRs, leveraging sequential clocking of gates. 
\item We investigate the performance of the proposed address decoder in terms of delay and power efficiency, using analysis and simulations. 
\begin{itemize}
\item We propose a novel architecture to obtain the maximum possible power efficiency. It can decode 32-bit addresses by consuming only 2nW at 1Kbps, which shows a significant improvement in power efficiency than the legacy architecture consuming roughly 12$\mu$W to decode a 16-bit address at the same bit rate \cite{ref12}.
\item The proposed architecture benefits a light strategy for dealing with noise, and it can detect addresses even though some of the address bits might be corrupted. The technique utilized to achieve this goal is to detect and overlook a certain number of corrupted address bits to a certain extent. 
\item
The address length in the proposed architecture can be adjusted to any length shorter than its maximum capacity. This address length flexibility brings efficiency as a single address decoder embedded in the device could be used in different lengths, e.g., shorter length for personal area communications, and full length in wide-area communications, in order to save energy further.
\end{itemize}
    \item We investigate the energy consumption of uplink and downlink IoT communications with/without WuR versus different communications parameters. Find operation regions in which integration of WuR is beneficial, investigate challenges in such integration, and potential solutions for compensation of challenges. 
    
\end{itemize}

The remainder of the paper is organized as follows. The system model and the problem description are given in Section II. 
In Section III, the proposed address decoding scheme is presented. In Section IV, modeling of KPIs and device-level performance evaluations are presented. In Section V, the integration of WuRs in uplink/downlink communications is investigated. Finally, the concluding remarks are given in Section~VI.

\begin{figure}
	\centering
	\includegraphics[width=1\columnwidth]{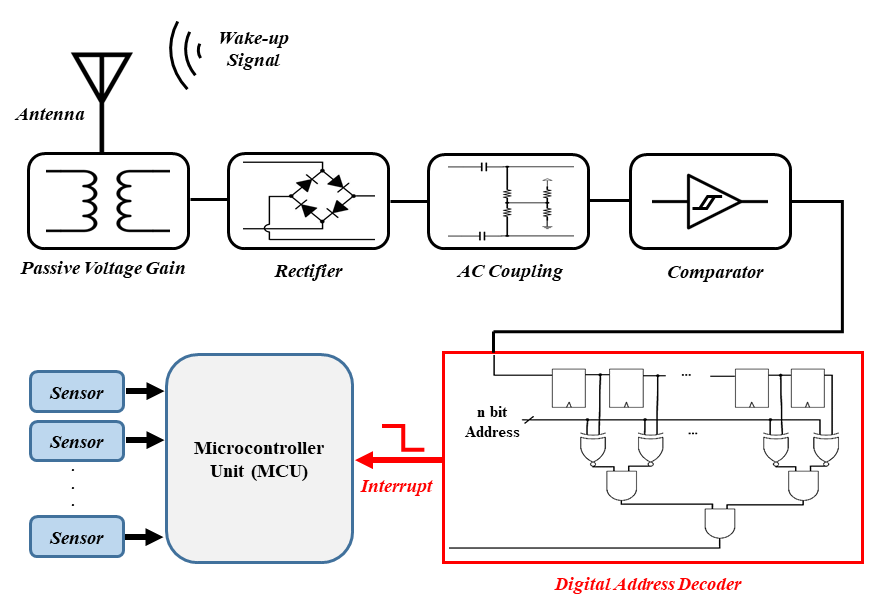}
	\caption{{A wake-up receiver unit.}}
	\label{fig1}
\end{figure} 

 \begin{figure*}
	\centering
	\includegraphics[width=5in]{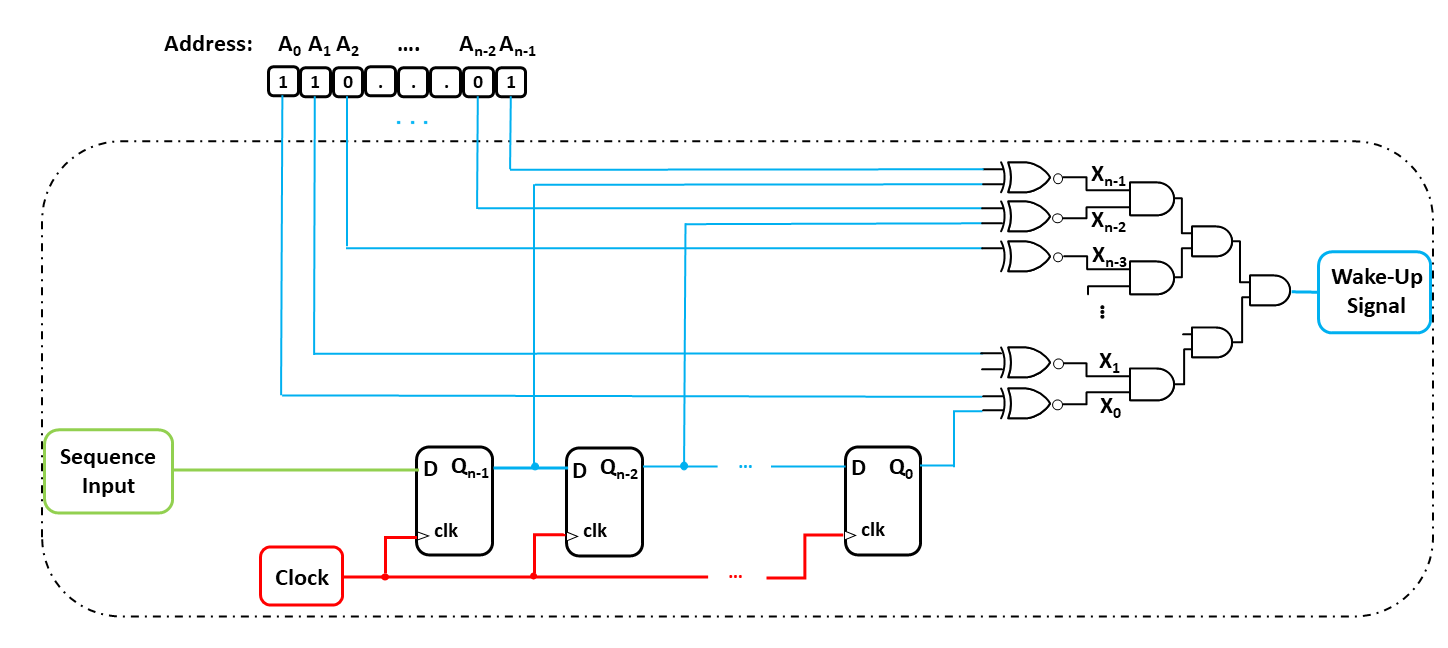}
	\caption{The legacy address decoder comprises a large network of XNOR gates as the address correlator, which consume a significant portion of dynamic power.} 
	\label{fig3}
\end{figure*}

 \section{System Model and Problem Description }\label{sys}
As illustrated in Fig. \ref{fig1}, a WuR consists of an analog demodulator and a digital address decoder. The analog demodulator comprises several modules, including an amplifier, rectifier, AC coupling, and comparator (the black boxes in Fig. \ref{fig1}). {The demodulated signal is then passed on toward a digital address decoding unit (the red solid area in Fig. \mbox{\ref{fig2}}) which checks if the received signal matches the saved address.} If the received signal matches the device's address, a wake-up interrupt is generated for the primary circuit.

\begin{figure*}[t!]
		\centering
		\includegraphics[ width=5in]{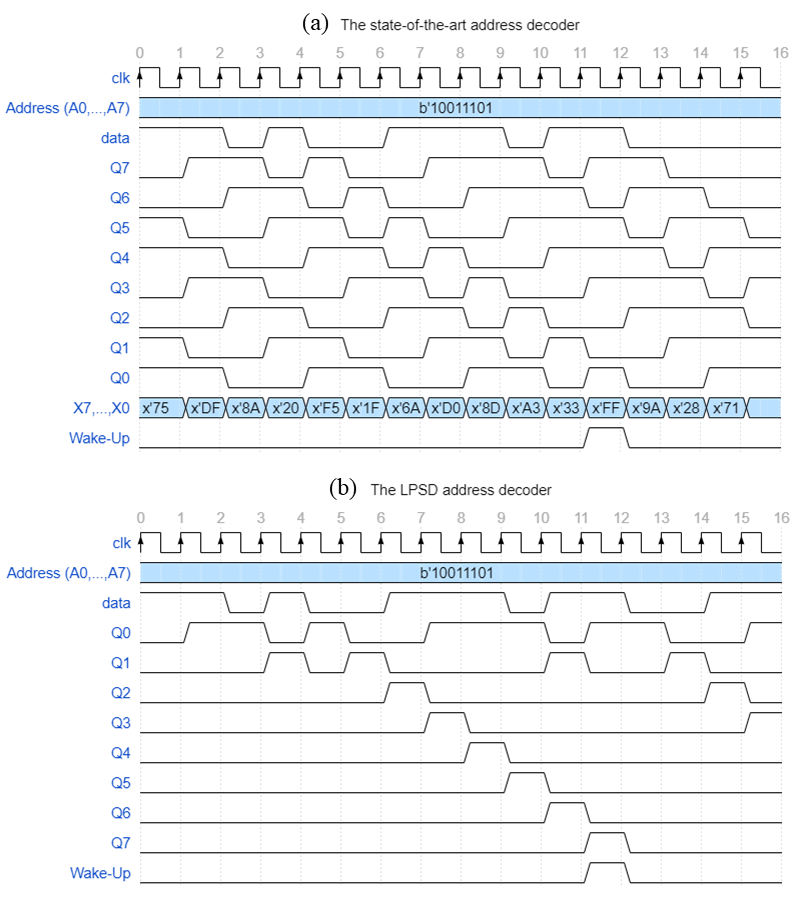}
		\caption{Transitions of the nodes in (a) the legacy and (b) the proposed LPSD address decoder for an arbitrary input and address (n=8).}
		\label{timing diagram}
	\end{figure*}

\begin{figure*}[t!]
		\centering
		\includegraphics[ width=5in]{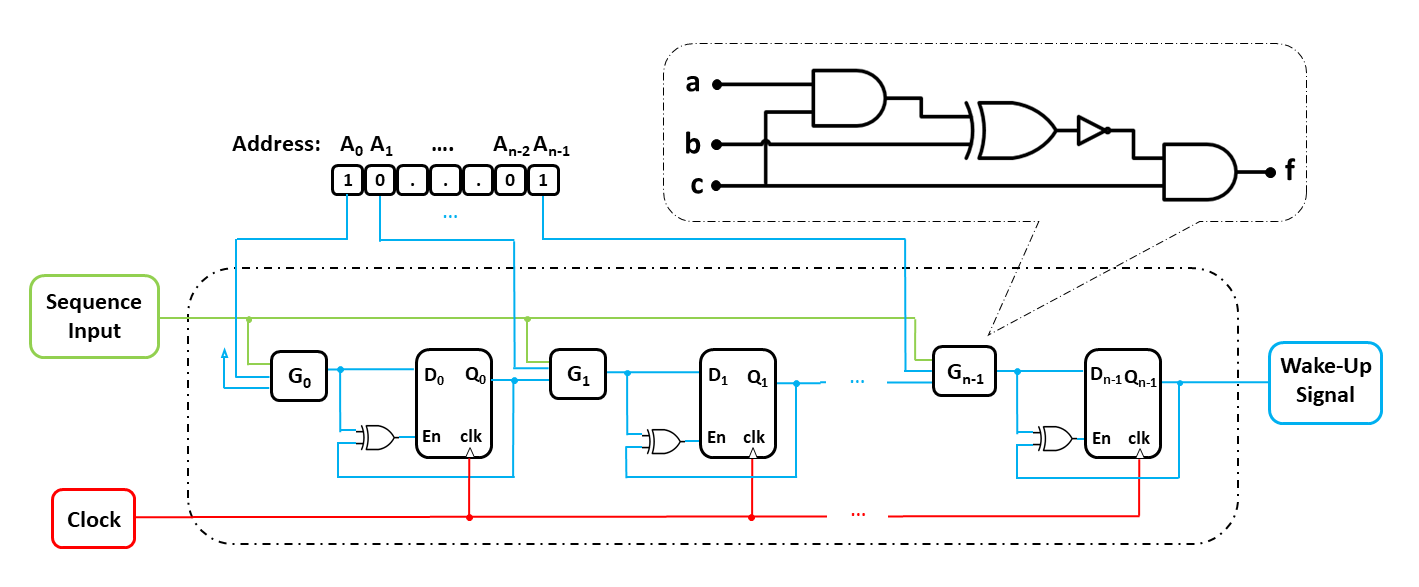}
		\caption{The LPSD's sequential architecture. Each stage consists of a G-block and a Flip-Flop. The G-block internal circuit is shown in the dashed box. }
		\label{fig2}
	\end{figure*}

The architecture of an $n$-bit address decoder has been depicted in  Fig. \ref{fig3} for a conventional WuR \cite{ref18,ref20,p14d}. In this architecture, the address decoding unit stores the latest received $n$ bits in a shift register, consisting 8 FFs, and passes the input sequence and the stored address through a tree-shaped structure of XNOR and AND gates for comparison. 
To better understand, the timing diagram for the operation of an 8-bit address decoder with such architecture is presented in Fig. \ref{timing diagram}-a. As one observes here, from top to bottom,  the  clock signal, the saved address in the device (10011101), the received sequence of data, 8 values of $Q_0$ to $Q_7$, related to the output of respective flip flops, 8 values of $X_0$ to $X_7$, related to the output of XNOR gates, and finally the wake-up signal have been depicted. 
The above-described address decoding method, empowered by XNORs,  has been vastly utilized in different WuR designs  \cite{ref18,ref20,p14d}. In order to save further energy for WuRs, it is of crucial importance to decrease the number of active components in the circuitry. Hence, having the XNOR gates and the FFs activated all the time for online comparison, a considerable amount of dynamic power is consumed even when the device has not been paged. This energy consumption has been visualized indirectly  in the  timing diagram of Fig. \ref{timing diagram}-a, where the output values of FFs  vary frequently. 
Then, the first  research question (RQ) could be expressed as follows:

{RQ.1}: Given an $n$-bit constant address and a dynamic bitstream (changes by the clock), how can one leverage a low-power digital circuit to carry the comparison out and generates a wake-up signal as soon as there is a match?
Now, assume we have the low-power comparator circuit. The next question comes from the reliability point of view. Consider the fact that there is a possibility of error in the reception of data, demodulation, and other stages. Then, it is vital to ensure our wake-up procedure is resilient to such errors. The first sub-RQ aims at addressing this concern, as follows.

{RQ.1.a}: How can one introduce $m$-bit error tolerance to the comparison procedure, i.e., if the received sequence matches the input sequence in more than $m-n$ positions, it generates a wake-up signal?   
Finally, let us consider a device with an $n$-bit address that has been deployed in an application where devices are limited in  number, and hence, a much shorter address suffices for their paging. The next sub-RQ aims to make the comparison flexible to enable energy-saving whenever possible, as follows.

{RQ.1.b}: How can one make the comparison procedure flexible to work with different lengths of the address in order to save energy whenever possible?

In the second research question, we aim to investigate operation regions (in terms of activity, i.e.,  the rate of reporting and paging) in which the introduction of WuRs to IoT devices can bring energy savings. Towards this end, we need to model energy saving in uplink/downlink IoT communications and see which kind of devices can benefit from WuRs. Then, this problem is expressed as follows.

{RQ.2}: Given the activity of IoT devices in terms of rates of paging  for downlink communications and reporting  for uplink communications, for which activity ranges, leveraging WuRs results in significant energy saving for IoT devices?

 In Section III, we present our solution for RQ.1. Then, in Section IV, we evaluate the performance of the proposed solution and answer RQ.1.a and RQ.1.b  Finally, in Section V, we answer RQ.2.

	 \begin{figure*}
	\centering
	\includegraphics[width=5.5in]{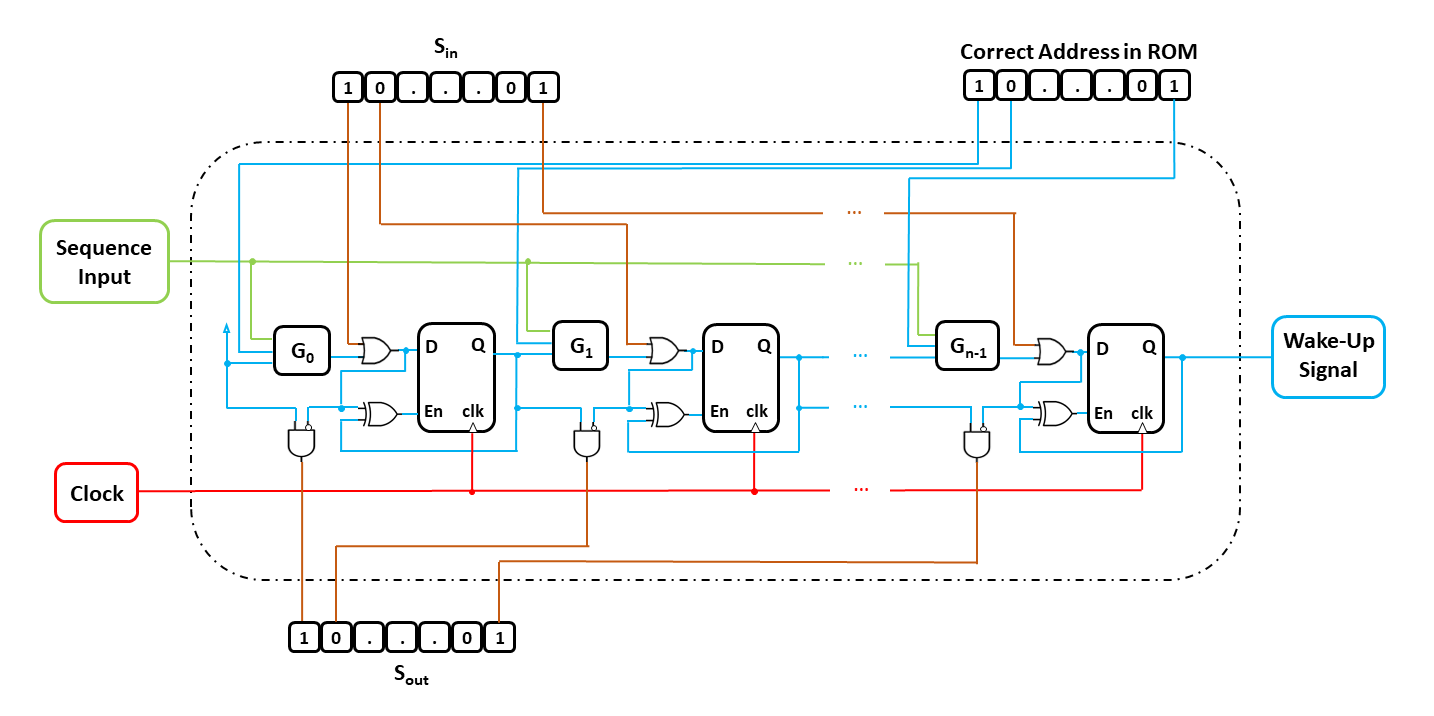}
	\caption{The $S_{in}$ and $S_{out}$ buses provide direct access to the LPSD's flip flips.} 
	\label{fig4}
\end{figure*}

 \begin{figure*}
	\centering
	\includegraphics[width=6in]{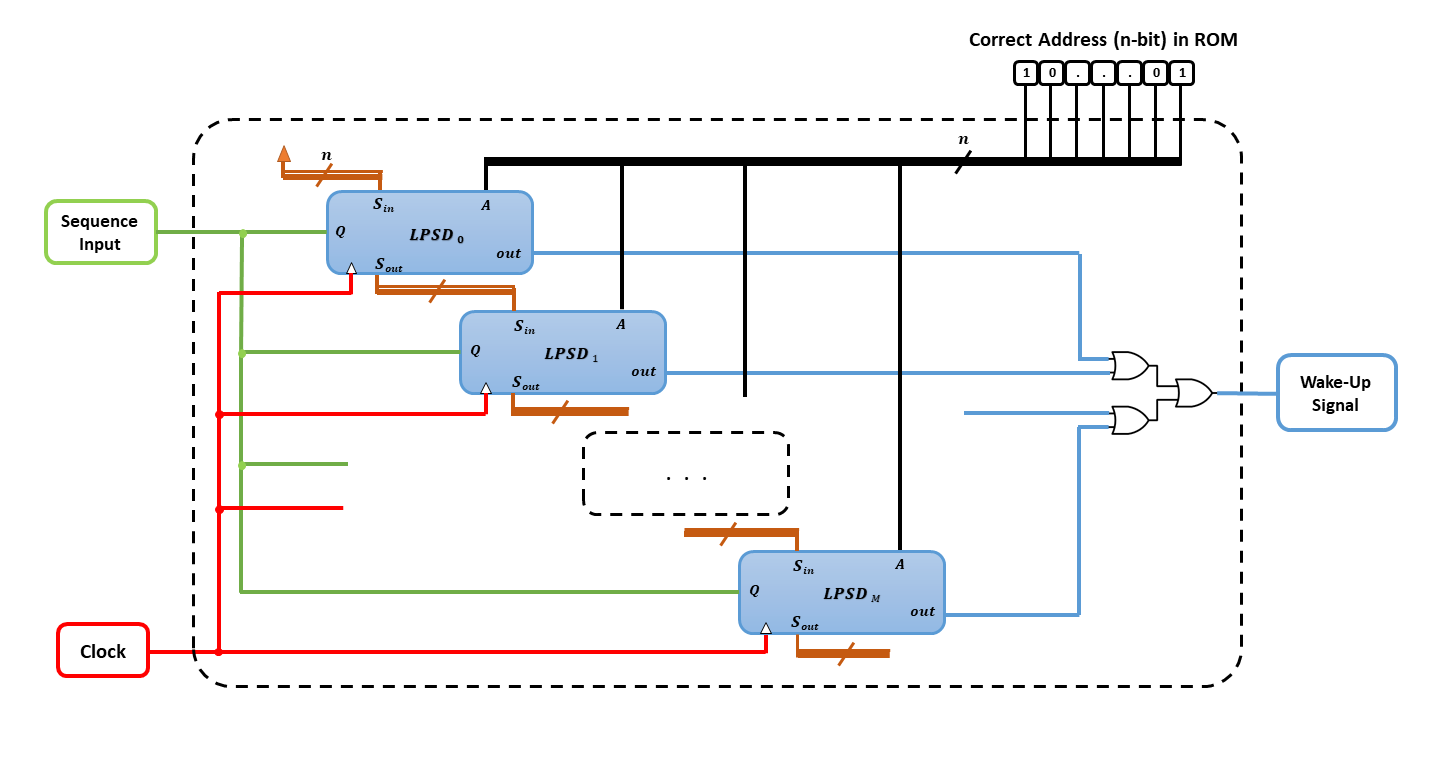}
	\caption{Redundant arrangement of LPSDs. Note that the sequence input, the correct address, and the wake-up signals are denoted as $Q$, $A$, and $W$ respectively.} 
	\label{fig5}
\end{figure*} 
\section{The Proposed Address Decoding Solution  }\label{SecSol}

This section aims to answer RQ.1, i.e., reducing the dynamic power consumption in address decoding. Recall from the timing diagram in Fig. \ref{timing diagram}-a, where there were many unnecessary transitions and activities in the comparator circuits before the wake-up signal is triggered.  Towards reducing power consumption in unnecessary transmissions, which are the primary sources of power consumption, we introduce  a low-power sequence decoder (LPSD) in which a sequential procedure for circuit-activity management is followed, as depicted in Fig. \ref{fig2}. The proposed LPSD leverages a novel block called G-blocks for bit-wise comparison and process flow control.  In this architecture, when a bit-match occurs, e.g., at the $A_0$ position of address as in Fig. \ref{fig2}, the respective G-block, i.e., $G_0$, approves it and stores its occurrence in the subsequent FF. The three inputs and the output of a G-block are denoted by $a$, $b$, $c$, and $f$, respectively (pop-outed section in Fig. \ref{fig2}). The ($a$,$b$,$c$) inputs are connected to the sequence input, the respective stored address bit in the memory (\(A_i\)), and the output state of the FF in the previous stage (\(Q_{i-1}\)), respectively. The $i^{th}$ G-block's output becomes 1 when \(Q_{i-1}\) is 1 (i.e., the latest $i$ bits have been correct so far), and the current input is also equal to \(A_i\). As Fig. \ref{timing diagram} and \ref{fig2} demonstrate, although the comparison employs an XOR gate in the G-block, it is not as active as the XNOR in the legacy design. Because in the legacy address decoder architecture (Fig. \ref{fig3}), one of the inputs of each XNOR gate is directly connected to the input data, which has a high transition probability. In contrast, in the proposed LPSD structure, an AND gate has been placed before the XOR gate to block propagation of the sequence input to the XOR gate. Hence, the XOR gate in the proposed solution  remains inactive primarily.

\begin{algorithm}
\caption{LPSD n-bit address decoder}\label{alg:1}
Reset all FFs\;
\While{\textup{true} }{
    Await for clock rising edge\;
   \Comment*[h]{\fontfamily{cmr}\selectfont{\small// Creating an array containing all of the set FFs indexes:}}\\
   $i\gets$ all of the set FF indexes\;
   \For{\textup{each} $i$ \textup{starting from the last}}{
   \If{$i$ \textup{is not the final index}}{
   \If{\textup{received bit is equal to} $address[i+1]$ }{
   Set $FF[i+1]$\;
   }
   Reset $FF[i]$\;
   }
   }
    \If{\textup{received bit is equal to} $address[0]$}{
    Set $FF[0]$\;
    }
   \If{\textup{the final FF is set }}{
   Trigger the wake-up signal\;
   } 
   Await for clock falling edge\; 
   
}
\end{algorithm}

{Regarding static power consumption, an XOR gate is connected to the En (enable) input of each FF, as shown in Fig.~\mbox{\ref{fig2}}.} This gate detects when a FF is likely to change its state on the clock's edges and enables it. It also deactivates the FF and puts it in sleep mode when no change in the FFs state will happen in the subsequent edge of the clock. From the address decoding perspective, the $i^{th}$ FF needs to change its stored bit in any of the following conditions:

\begin{itemize}
    \item 
    The transition from 0 to 1 is needed:
    \begin{itemize}
        \item the $i^{th}$ FF has been in state 0, \textbf{and}
        \item the last $i$ bits are  matching the corresponding part of the  stored address (i.e., the output of the $i^{th}$ G-block is 1)
     \end{itemize}   
     \item
     The transition from 1 to 0 is needed:   
     \begin{itemize}
    \item
    the $i^{th}$ FF is already set to 1 (because in the last previous clock cycle, all the last $i$ bits have matched the corresponding part of the stored address), \textbf{and}
    \item
    The latest $i$ bits are not matching anymore. 
    \end{itemize}
\end{itemize}

{Based on our design, if any of the above conditions are true, the XOR gate's output will become 1, and hence, the $i^{th}$ FF will be activated to apply the changes, with the exception of the first FF which is always active. The LPSD procedure can be better understood through Algorithm~\mbox{\ref{alg:1}}}. As the algorithm suggests, all of the LPSD’s FFs are initially reset. To sync the procedure with the {clock signal, the main procedure is wrapped between two await commands, checking for rising and falling edges. The array} {$i$, contains the indexes of every set FF. Using the indexes stored in $i$, the algorithm can determine that in which stages the comparison must carry; except for the first stage which must always do the comparison regardless of the other FFs status and hence it is separately written in the algorithm. Ultimately, the last if statement checks the status of the last FF for triggering the wakeup signal. In other words, the circuit design paradigm has been shifted from always-on to always available, i.e., any module is activated once needed.}
In the legacy address validation approaches, as depicted in Fig. \ref{timing diagram}, the entire received sequence content is in operation in every clock cycle \cite{ref20}. Hence, the wake-up interrupt command is produced when all  XNOR gates' outputs are 1 (in clock cycle 11 in Fig. \ref{timing diagram}-a). However, the proposed LPSD works sequentially, and hence, the majority of the LPSD circuit is generally in the inactive mode. {As the chance of receiving a 1 or 0 bit in a pure noise ambient is equal, the probability of random activation of the first FF of LPSD, i.e., $\text{FF}_0$ in Fig. 5, is  50\%. Then in the subsequent FF, the probability of random activation is half of  its predecessor. For example, the probability of random activation of the fifth FF, $\text{FF}_4$, by ambient noise is ${\frac{1}{2^5}\approx3.1\%}$.} Thus, most of LPSD's logic circuit remains inactive in practice, as depicted in the timing diagram of Fig.~\ref{timing diagram}-b. {From this figure, it is seen that the state of the fourth and higher FFs are unlikely to change their state.} Thereby, the device consumes a negligible amount of  power. While comparing the timing diagrams of the legacy and the proposed LPSDs in Fig. \ref{timing diagram} clearly shows the superior performance of the proposed scheme in power consumption, we will carry out an in-depth performance analysis in the next section.

\section{KPIs and Performance Evaluation at the Decoder Level}
In this section, we investigate the key performance indicators (KPIs) of the proposed address decoder. The KPIs of interest include the reliability of address decoding, latency in address decoding, power consumption in continuous sequence decoding, and solution flexibility to address length. Furthermore, we answer RQ.1.a (presenting an extended version of the LPSD, which is tolerating mismatch in address decoding) and RQ.1.b (flexibility of LPSD to address length) in this section.

\subsection{Reliability}\label{reli}
Data corruption due to environmental noise or interference is a fundamental limitation of any wireless communication system and must be considered in the design and operation management of its entities and the services offered through it. {To have a probabilistic analysis of this limitation in our problem, let us denote by $p_b$ the bit-error rate (BER). In \mbox{\cite{m14p}}, $p_b$ has derived for an OOK address detector operating in additive white Gaussian noise (AWGN) channel, as: $p_b=c e^{-\lambda}.$ In this expression,  $\lambda$ is the signal-to-noise ratio (SNR) of the envelope detector's input, and $c$ is a constant.} A consecutive sequence of correct address bits must be received for a conventional address decoder, such as the one leveraged in {\mbox{\cite{m14p}}}, to trigger a wake-up signal. Hence, a single corruption in the sequence traps the full detection procedure, and no signal will be triggered until the full address is re-transmitted. Then, the probability of a conventional $n$-bit address decoder detecting an address correctly is given by
{$P_{\text{conv}}(n)=(1-p_b)^n$.}

In order  to avoid being trapped by less than $m$ bits corrupted bits in address decoding, we proposed to follow a 2-step procedure: (i) allocate address codes such that a minimum hamming distance of $2m+1$ is achieved between any two devices\footnote{This is to make sure tolerance to mismatch in address decoding does not result in the generation of false wake-up signals.}, and (ii) design the address decoder such that it accepts $m$ mismatches between received sequence and the store address. Recall from the description of the proposed LPSD in Fig. \ref{fig2} that investigation of the received sequence proceeds as the first received bit matches the first bit of the address and stops as soon as a mismatch occurs. To count and ignore a certain number of mismatches, the presented architecture for LPSD in Fig. \ref{fig2} is improved by adding $S_{in}$ and $S_{out}$ buses, as shown in Fig. \ref{fig4}.  Then,  multiple redundant LPSD blocks are cascaded in parallel. When a mismatch occurs, the procedure stops at the corresponding LPSD and is continued in the adjacent LPSD,  as shown in the improved architecture of  Fig. \ref{fig5}. Then, to bring tolerance to $m$ bits of mismatch, $m$ adjacent LPSDs are needed in parallel.
\newcommand{\mathcolorbox}[2]{\colorbox{#1}{$\displaystyle #2$}}
Following the derivation of $p_b$ in the above, the probability of correct detection of an $n$-bit address by using the updated architecture for tolerating $m$-bit mismatch is derived as:
\begin{equation}\mathcolorbox{yellow}{P_{\text{LPSD}(n,m)}=\sum_{k=0}^{m}\left(\genfrac{}{}{0pt}{}{n}{k}\right)(1-p_b)^{n-k}(p_b)^k\label{eq3}}\end{equation}
Comparison of $P_{\text{LPSD}}(n,m)$ with $P_{\text{conv}}(n)$ reveals that since for a non-zero $m$,
$P_{\text{LPSD}}(n,m)>P_{\text{conv}}(n)$, and the difference becomes much higher by an increase in $m$. Then, by choosing an appropriate value of $m$, in accordance with the environment, one can minimize the probability of missing a wake-up signal.

\subsection{Latency Analysis}

\begin{figure}[!t]
	\centering
	\resizebox{.95\columnwidth}{!}{
%
%
\definecolor{mycolor1}{rgb}{0.00000,0.44700,0.74100}%
\definecolor{mycolor2}{rgb}{0.85000,0.32500,0.09800}%
\definecolor{mycolor3}{rgb}{0.92900,0.69400,0.12500}%
\begin{tikzpicture}

\begin{axis}[%
width=6.028in,
height=4.754in,
at={(1.011in,0.642in)},
scale only axis,
bar shift auto,
bar width=8mm,
symbolic x coords={8-bit,16-bit,32-bit,64-bit},
enlarge x limits=0.2,
xtick=data,
xlabel style={font=\color{white!15!black}},
xlabel={\LARGE Address Length},
ymin=0,
ymax=12,
xmajorgrids,
xminorgrids,
ymajorgrids,
yminorgrids,
ylabel style={font=\color{white!15!black}},
ylabel={\LARGE Delay (ns)},
axis background/.style={fill=white},
legend style={legend cell align=left, align=left, draw=white!15!black}
]
\addplot[ybar, fill=mycolor1, draw=black, area legend] table[row sep=crcr] {%
2-bit 0\\
8-bit	1.1\\
16-bit	2.1\\
32-bit	4.6\\
64-bit	8.1\\
};

\addlegendentry{ \LARGE  State-of-the-Art}

\addplot[ybar, fill=mycolor2, draw=black, area legend] table[row sep=crcr] {%
2-bit 0\\
8-bit	1.3\\
16-bit	2.3\\
32-bit	4.8\\
64-bit	8.3\\
};
\addlegendentry{ \LARGE  LPSD - 1-bit corruption robustness}

\addplot[ybar, fill=mycolor3, draw=black, area legend] table[row sep=crcr] {%
2-bit 0\\
8-bit	1.5\\
16-bit	2.5\\
32-bit	5\\
64-bit	8.5\\
};

\addlegendentry{ \LARGE  LPSD - 2-bit corruption robustness}

\end{axis}

\end{tikzpicture}%
}
	\caption{LPSD's delay in different address lengths} 
	\label{fig7}
\end{figure}
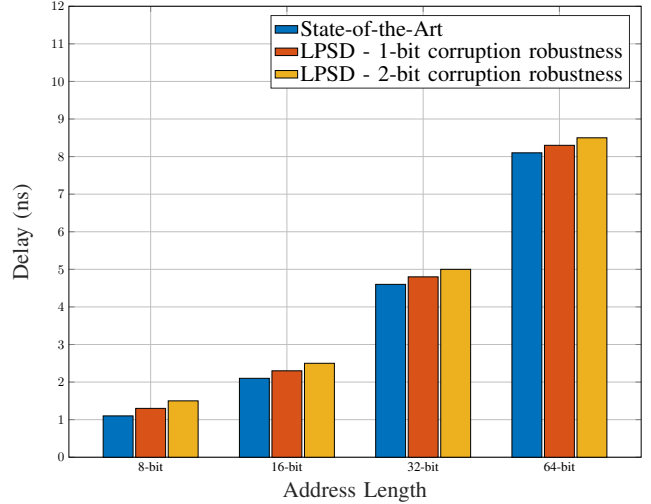

One of the vital statistical aspects of every digital device is the minimum time at which the device can propagate its input signals towards the outputs. This time length is essentially the minimum required time that the device needs to execute a single operation and is  determined by finding the longest path that an input signal needs to pass to reach the output terminals.
The maximum time delay of the legacy and proposed $n$-bit address decoders  are derived as follows:
\begin{align}
\hat{T}_{\text{conv}}&=n T_{\text{FF}_{t}}+(\log_2 n +1)T_{\text{Gate}}\label{eq4}\\
\hat{T}_{\text{LPSD}}&=n (T_{\text{FF}_{t}}+T_{\text{FF}_\text{{en}}}+2T_{\text{Gate}})+T_{G},\label{eq5}
\end{align}
in which, $\hat{T}$, $T_{\text{FF}_{t}}$, $T_{\text{FF}_\text{{en}}}$, $T_{\text{Gate}}$, and $T_{G}$ denote the maximum time delay of the device, the toggle and enable delay values of an FF, the logic gate delay, and the time delay of a G-block, respectively. From the G-block description in Fig. \ref{fig2}, $T_{G}$  is equal to $4T_{\text{Gate}}$, respectively.
From digital logic design \cite{navabi1999verilog}, $T_{\text{Gate}}<T_{\text{\text{FF}}_\text{{en}}}\ll T_{\text{\text{FF}}_{t}}$. Hence, for large values of $n$, the dominant element is $T_{\text{\text{FF}}_{t}}$, and the rest could be neglected.  Thus, \eqref{eq4}-\eqref{eq5} could be simplified as:
\begin{align}
\hat{T}_{\text{conv}}\approx n T_{\text{\text{FF}}_{t}}+t_1,\label{eq6}\\
\hat{T}_{\text{LPSD}}\approx n T_{\text{\text{FF}}_{t}}+t_2,\label{eq7}
\end{align}
in which $t_1$ and $t_2$ are constant time delay values, and $t_1<t_2$. From the LPSD architecture in Fig. \ref{fig2}, one observes that adding $m$ redundant stages of LPSD for error tolerance merely adds $m\times T_{\text{Gate}}$ delay to $\hat{T}_{\text{LPSD}}$, adding the potential of recovery from bit corruption to the address decoder. One must note that this extra latency for noise robustness is negligible for large values of $n$, i.e., the address length, as one observes in Fig. \ref{fig7}. This figure depicts the maximum delay of the legacy and proposed LPSD architectures for different address lengths. From these results, one observes that delay performances of both architectures are roughly the same, and they follow a trend in order of $n$ in bytes. In other words, the maximum delay  for an $n$-bit address decoder is approximately $n/8$ nano-seconds. 


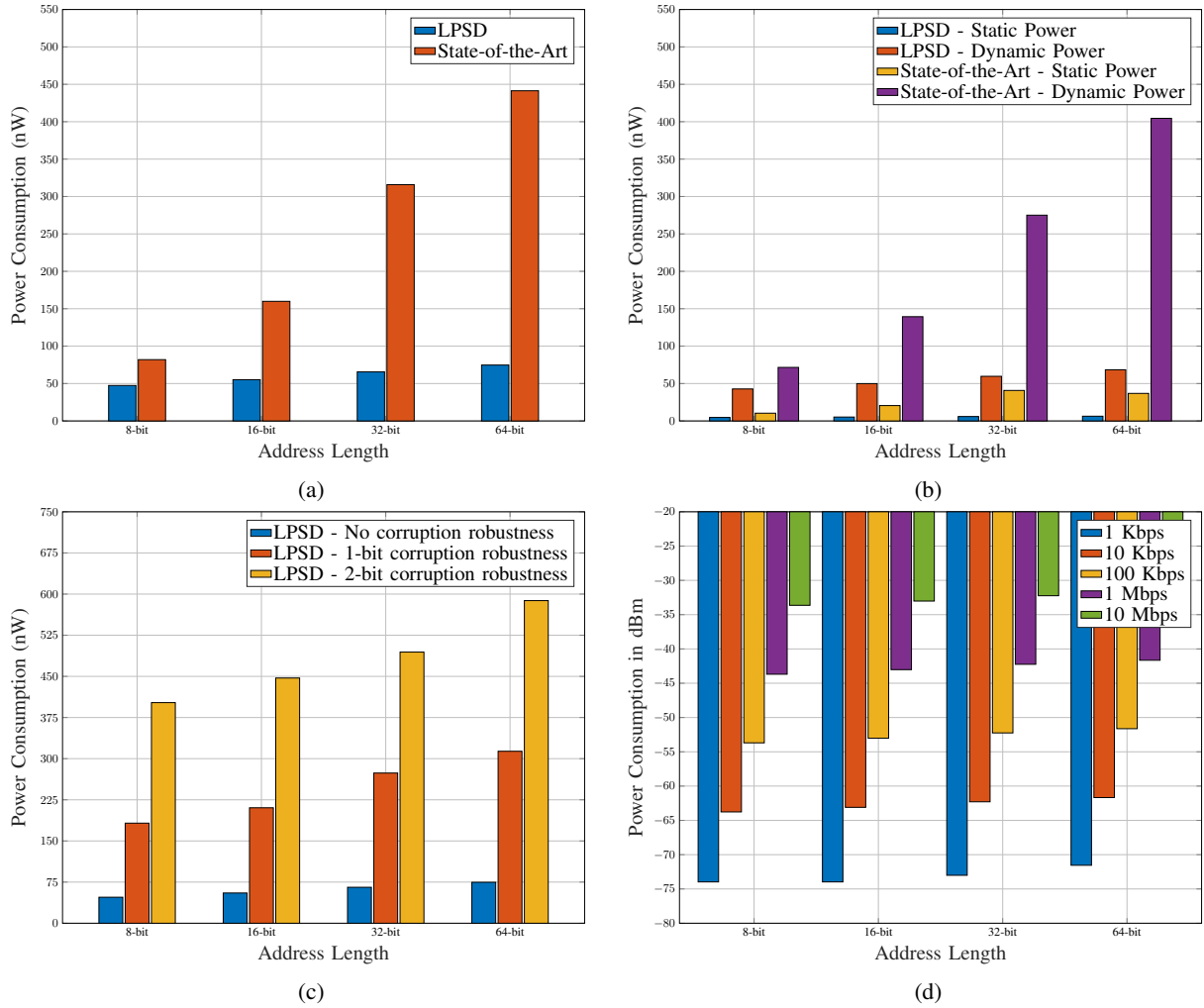
\begin{figure*}
    \centering
    \begin{subfigure}[t]{.45\textwidth}
        \resizebox{.95\columnwidth}{!}{
%
%
\definecolor{mycolor1}{rgb}{0.00000,0.44700,0.74100}%
\definecolor{mycolor2}{rgb}{0.85000,0.32500,0.09800}%
\begin{tikzpicture}

\begin{axis}[%
width=6.028in,
height=4.754in,
at={(1.011in,0.642in)},
scale only axis,
bar shift auto,
bar width=8mm,
symbolic x coords={8-bit,16-bit,32-bit,64-bit},
enlarge x limits=0.2,
xtick=data,
xlabel style={font=\color{white!15!black}},
xlabel={\LARGE Address Length},
ymin=0,
ymax=550,
xmajorgrids,
xminorgrids,
ymajorgrids,
yminorgrids,
ylabel style={font=\color{white!15!black}},
ylabel={\LARGE Power Consumption (nW)},
axis background/.style={fill=white},
legend style={legend cell align=left, align=left, draw=white!15!black}
]
\addplot[ybar, fill=mycolor1, draw=black, area legend] table[row sep=crcr] {%
2-bit	0\\
8-bit	47.42\\
16-bit	55.12\\
32-bit	65.56\\
64-bit	74.69\\
};

\addlegendentry{\LARGE LPSD}

\addplot[ybar, fill=mycolor2, draw=black, area legend] table[row sep=crcr] {%
2-bit	0\\
8-bit	81.95\\
16-bit	159.91\\
32-bit	315.85\\
64-bit	441.35\\
};

\addlegendentry{\LARGE State-of-the-Art}

\end{axis}

\end{tikzpicture}%
}
        \caption{}
        \label{fig:my_label1}
     \end{subfigure}
     \begin{subfigure}[t]{.45\textwidth}
        \resizebox{.95\columnwidth}{!}{
%
%
\definecolor{mycolor1}{rgb}{0.00000,0.44700,0.74100}%
\definecolor{mycolor2}{rgb}{0.85000,0.32500,0.09800}%
\definecolor{mycolor3}{rgb}{0.92900,0.69400,0.12500}%
\definecolor{mycolor4}{rgb}{0.49400,0.18400,0.55600}%
\begin{tikzpicture}

\begin{axis}[%
width=6.028in,
height=4.754in,
at={(1.011in,0.642in)},
scale only axis,
bar width=6mm,
bar shift auto,
symbolic x coords={8-bit,16-bit,32-bit,64-bit},
enlarge x limits=0.2,
xtick=data,
xlabel style={font=\color{white!15!black}},
xlabel={ \LARGE Address Length},
ymin=0,
ymax=550,
xmajorgrids,
xminorgrids,
ymajorgrids,
yminorgrids,
ylabel style={font=\color{white!15!black}},
ylabel={\LARGE Power Consumption (nW)},
axis background/.style={fill=white},
legend style={legend cell align=left, align=left, draw=white!15!black}
]
\addplot[ybar, fill=mycolor1, draw=black, area legend] table[row sep=crcr] {%
2-bit 0\\
8-bit	4.66\\
16-bit	5.22\\
32-bit	5.89\\
64-bit	6.33\\
};

\addlegendentry{ \LARGE  LPSD - Static Power}

\addplot[ybar, fill=mycolor2, draw=black, area legend] table[row sep=crcr] {%
2-bit 0\\
8-bit	42.76\\
16-bit	49.9\\
32-bit	59.67\\
64-bit	68.36\\
};

\addlegendentry{ \LARGE  LPSD - Dynamic Power}

\addplot[ybar, fill=mycolor3, draw=black, area legend] table[row sep=crcr] {%
2-bit 0\\
8-bit	10.38\\
16-bit	20.56\\
32-bit	40.78\\
64-bit	36.83\\
};

\addlegendentry{ \LARGE  State-of-the-Art - Static Power}

\addplot[ybar, fill=mycolor4, draw=black, area legend] table[row sep=crcr] {%
2-bit 0\\
8-bit	71.57\\
16-bit	139.35\\
32-bit	275.07\\
64-bit	404.52\\
};

\addlegendentry{ \LARGE  State-of-the-Art - Dynamic Power}

\end{axis}

\end{tikzpicture}%
}
        \caption{}
        \label{fig:my_label2}
     \end{subfigure}
    \begin{subfigure}[t]{.45\textwidth}
        \resizebox{.95\columnwidth}{!}{
%
%
\definecolor{mycolor1}{rgb}{0.00000,0.44700,0.74100}%
\definecolor{mycolor2}{rgb}{0.85000,0.32500,0.09800}%
\definecolor{mycolor3}{rgb}{0.92900,0.69400,0.12500}%
\begin{tikzpicture}

\begin{axis}[%
width=6.028in,
height=4.754in,
at={(1.011in,0.642in)},
scale only axis,
bar shift auto,
bar width=7mm,
symbolic x coords={8-bit,16-bit,32-bit,64-bit},
enlarge x limits=0.2,
xtick=data,
xlabel style={font=\color{white!15!black}},
xlabel={\LARGE  Address Length},
ymin=0,
ymax=750,
ytick distance={75},
xmajorgrids,
xminorgrids,
ymajorgrids,
yminorgrids,
ylabel style={font=\color{white!15!black}},
ylabel={\LARGE  Power Consumption (nW)},
axis background/.style={fill=white},
legend style={legend cell align=left, align=left, draw=white!15!black}
]
\addplot[ybar, fill=mycolor1, draw=black, area legend] table[row sep=crcr] {%
2-bit   0\\
8-bit	47.42\\
16-bit	55.12\\
32-bit	65.56\\
64-bit	74.69\\
};
\addlegendentry{ \LARGE LPSD - No corruption robustness}

\addplot[ybar, fill=mycolor2, draw=black, area legend] table[row sep=crcr] {%
2-bit   0\\
8-bit	182.46\\
16-bit	210.5\\
32-bit	273.77\\
64-bit	313.49\\
};

\addlegendentry{ \LARGE LPSD - 1-bit corruption robustness}

\addplot[ybar, fill=mycolor3, draw=black, area legend] table[row sep=crcr] {%
2-bit   0\\
8-bit	402.15\\
16-bit	447\\
32-bit	494.29\\
64-bit	588.11\\
};

\addlegendentry{ \LARGE LPSD - 2-bit corruption robustness}

\end{axis}

\end{tikzpicture}%
}
        \caption{}
        \label{fig:my_label3}
     \end{subfigure}
    \begin{subfigure}[t]{.45\textwidth}
        \resizebox{.95\columnwidth}{!}{
%
%
\definecolor{mycolor1}{rgb}{0.00000,0.44700,0.74100}%
\definecolor{mycolor2}{rgb}{0.85000,0.32500,0.09800}%
\definecolor{mycolor3}{rgb}{0.92900,0.69400,0.12500}%
\definecolor{mycolor4}{rgb}{0.49400,0.18400,0.55600}%
\definecolor{mycolor5}{rgb}{0.46600,0.67400,0.18800}%
\begin{tikzpicture}

\begin{axis}[%
width=6.028in,
height=4.754in,
at={(1.011in,0.642in)},
scale only axis,
bar shift auto,
bar width=6mm,
symbolic x coords={ 8-bit,16-bit,32-bit,64-bit},
enlarge x limits=0.2,
xtick=data,
xlabel style={font=\color{white!15!black}},
xlabel={\LARGE Address Length},
ymin=-80,
ymax=-20,
ylabel style={font=\color{white!15!black}},
ylabel={\LARGE Power Consumption in dBm},
axis background/.style={fill=white},
xmajorgrids,
xminorgrids,
ymajorgrids,
yminorgrids,
legend style={legend cell align=left, align=left, draw=white!15!black}
]
\addplot[ybar, fill=mycolor1, draw=black, area legend] table[row sep=crcr] {%
2-bit   0\\
8-bit	-73.9750\\
16-bit	-73.9750\\
32-bit	-73.0100\\
64-bit	-71.5490\\
};
\addlegendentry{\LARGE 1 Kbps}

\addplot[ybar, fill=mycolor2, draw=black, area legend] table[row sep=crcr] {%
2-bit   0\\
8-bit	  -63.7675\\
16-bit	-63.0980\\
32-bit	-62.2914\\
64-bit	 -61.6749\\
};
\addlegendentry{\LARGE 10 Kbps}

\addplot[ybar, fill=mycolor3, draw=black, area legend] table[row sep=crcr] {%
2-bit   0\\
8-bit	-53.6957\\
16-bit	-53.0190\\
32-bit	-52.2475\\
64-bit	-51.6304\\
};
\addlegendentry{\LARGE 100 Kbps}

\addplot[ybar, fill=mycolor4, draw=black, area legend] table[row sep=\\] {%
2-bit   0\\
8-bit	-43.6896\\
16-bit	-43.0190\\
32-bit	-42.2424\\
64-bit	-41.6520\\
};
\addlegendentry{\LARGE 1 Mbps}

\addplot[ybar, fill=mycolor5, draw=black, area legend] table[row sep=\\] {%
2-bit   0\\
8-bit	-33.6375\\
16-bit	-33.0184\\
32-bit	-32.2418\\
64-bit	-31.6581\\
};
\addlegendentry{\LARGE 10 Mbps}

\end{axis}

\end{tikzpicture}%
}
        \caption{}
        \label{fig:my_label4}
     \end{subfigure}
     \caption{Power consumption comparison between LPSD and	the legacy architecture in different address lengths: (a) Total power consumption comparison  at 1Mbps; (b) Dynamic and static power consumption at 1Mbps; (c) Impact of adding noise robustness modules to the design on power consumption, (d) Impact of the bit rate of address decoder on power consumption.}
     \label{fig6}

\end{figure*}

\subsection{Power Consumption Evaluation}

In this section, we investigate the power consumption of the address decoder module as the main KPI of interest and compare it against the legacy solutions. The following results, in Fig. \ref{fig6}, have been obtained by executing several rounds of power analyses under different configurations, including address length and bitrate (the rate at which, address decoder is operating). The simulations have been executed in Synopsys Design Compiler with UMC65 library, powerful software for detailed digital device analysis. {A sequence of random input data was generated (Length=5000) for this purpose while some correct address sequences were included in it. To be clearer, we manipulated this sequence manually according to the presumed address so that 7\%, 10\% and 20\% of the bits in the sequence were a part of a totally correct address, an approximately correct address with one or two error bits, and a half correct address, respectively. The rest of the bits remained random and unchanged. Obviously, in a condition that all the address bits were purely random the resultant power consumption of the proposed structure could have been even higher in view of lower activity of the FFs. This procedure was carried out the same for all of the structures in this work.}
 
 Now, let us take a deeper look at the power consumption evaluation results.  Fig. \ref{fig:my_label1} compares the total power consumption of the proposed LPSD against the legacy architecture \cite{ref18,ref20,p14d}.  {One observes  that the consumed  power by the legacy solution is roughly 6 times more than the power consumption of the proposed approach for a 64-bit address.} Even for a short address length of 16 bits, we observe that the proposed architecture outperforms the legacy approach by reducing the average power consumption in continuous decoding by 67\%, i.e., the legacy approach consumes 3 times more power. One must note that this level of reduction has been made possible by detecting and removing the primary sources of power consumption, i.e., unnecessary activity of different parts of the comparator circuit. A detailed analysis of power consumption, including specifying  dynamic and static 
parts of power consumption, has been presented in Fig. \ref{fig:my_label2}.
 The analytics in this figure suggests dynamic power forms roughly over 95\% of the overall device power consumption in both the legacy and LPSD architectures. However, although static power forms an insignificant portion of the overall power, it drains continuously from the power source. Hence, it is also a significant issue, especially when the device is expected to be in the standby state for a long time. From the simulation results in  Fig. \ref{fig:my_label2}, one observes that in LPSD architecture, both static and dynamic power is reduced significantly compared to the legacy approach. From Fig. \ref{fig:my_label1} and Fig. \ref{fig:my_label2}, we further notice that power consumption of the legacy approach increases almost linearly with the length of address, while for the proposed LPSD, the change in power consumption by doubling the address length is less than 20\%. {To be more specific, for the legacy and proposed approaches, by increasing the address length eight times, the power consumption has increased 440\% and 57\%, respectively.}

Fig. \ref{fig:my_label2} reflects the impact of adding modules for noise robustness on the power consumption. As discussed in Section \ref{reli}, multiple LPSDs can operate redundantly so that the address decoder can ignore a certain number of corrupted bits. However, adding such redundant LPSDs, increases the number of  active modules in the WuR, and consequently, the overall power  might increase. Since the noise robustness level depends on the number of redundant LPSDs one observes a direct trade-off between power efficiency and the level of robustness. 
The results show that a 64-bit address decoder based on the proposed architecture and 1-bit robustness consumes 30\% less power than the legacy approach without corruption robustness.

{Finally, in Fig.~\mbox{\ref{fig:my_label4}} we investigate the impact of the bit rate of address decoding on power consumption. In this figure (where power consumption is reported in dBm due to its logarithmic profile, hence a wide range of bit rates could be distinguished), we observe that the impact of address length on the power consumption (the $x$-axis) is negligible compared to the impact of frequency of operation. Furthermore, it is seen that increasing the bit rate of operation by 10 times results in an approximately 10dB increase in power consumption.}

{Table \mbox{\ref{tabarea}} demonstrates that although the LPSD architecture has been quite beneficial, it roughly occupies up to 60\% more area than the state-of-the-art, which can be considered the cost of this architecture. As an example, for 64-bit address length, the LPSD requires about 1900 $\mu m^2$ to implement using 65nm CMOS technology while the state-of-the-art takes about 800 $\mu m^2$ less with the same technology. Furthermore, adding redundant LPSD layers for error resiliency can also increase the extent of the area.}

 \begin{table}
    \centering
	\caption{{Different address decoder architecture implementation areas.}}
	\setlength{\tabcolsep}{3pt}
		\begin{tabular}{p{2 cm}p{1.1 cm}p{1.2 cm}p{1.2 cm}p{1.2 cm}}\\
		\toprule[0.5mm]
		\multicolumn{5}{c}{Area ($\mu$m$^2$)}\\[1ex] \hline\hline
		Address length &  8-bit &  16-bit &  32-bit &  64-bit \\[2ex] \hline
		State-of-the-Art &  137.88 & 279 & 561.6 & 1126.6  \\[2ex]
		LPSD &  191.88 & 430.92 & 909.36 & 1866.24 \\[2ex]
		LPSD ($m=1$) &  528.48 & 1156.32 & 2366.64 & 4925.52  \\[2ex]
		LPSD ($m=2$) &  794.52 & 1736.28 & 3551.76  & 7390.08 \\[1ex] \hline
     
	\end{tabular}
	\label{tabarea}
\end{table}

\begin{table}
\centering
	\caption{A comparison of different address decoder technologies. {$ \eta $ (\%) is the power consumption ratio of the address decoder to WuR.}}
	\setlength{\tabcolsep}{3pt}
		\begin{tabular}{p{1 cm}p{1 cm}p{1 cm}p{1 cm}p{1 cm}p{1 cm}p{1 cm}}\\
			\toprule[0.5mm]
      Ref &  Tech. &  Power ($\mu$W) &  Bit Rate (Kbps) &   Voltage (V) &  Address (bit) & {$\eta$ (\%)} \\[1ex] \hline\hline
     {This work} &  LPSD & 0.068 & 1000 & 0.9 & 64 & -   \\[3ex]
     \cite{ref18} &  FF & 13.14 & 100 & 0.9 & 16 & 97.9  \\[3ex]
     \cite{ref20} &  FF & 25.5 & 1 & 3 & 16 & 65.3   \\[3ex]
     \cite{p14d} &  FF & 101.7 & - & 0.9 & 18 & 34.64 \\[3ex]
     \cite{ref22} &  MCU & 70.6 & 500 & 0.8 & 8 & 98 \\[0ex] \hline
     
	\end{tabular}
	\label{tab1}
\end{table}

\subsection{Address Length Flexibility}

As mentioned earlier, the proposed LPSD can adjust its address to shorter lengths as well. The address length can be shrunk by bypassing a certain number of LPSD’s stages and leaving only the required number of them functional. Bypassing the stages is feasible using the $S_{in}$ bus, which provides direct access to write on each stage's FF. Since each FF stores the comparison result of that stage, pulled-up FFs do not require matching approval to propagate the sequence, and hence, they do not affect the procedure. Then, by leveraging the design in Fig. \ref{fig4}, we have an address decoder that enables operation with a flexible address length (response to RQ.1.c). 

A summary of performance  evaluation results has been presented in Table~\ref{tab1}. According to this table, the proposed architecture consumes much less power than the most efficient low-power address decoding technology we found in the literature, which is an incredible improvement in the address decoder's power efficiency. {In Table~\mbox{\ref{tab1}}, only the power consumption of the address decoder part of WuR for the various works is examined.} Furthermore, the proposed design operates twice as fast as its fastest prior art \cite{ref22}, and its supply voltage is amongst the lowest. {Also, to check the importance of the address decoder part in the WuRs, we define a parameter ($\eta$) to indicate what percentage of the power consumption of the WuR is due to the address decoder part. The value of $\eta$ for different works is given in Table~\mbox{\ref{tab1}}. We can say that the address decoder consumes most of the power of the WuR.}

 \begin{figure*}[t!]
    \centering
    \begin{subfigure}[t]{1\textwidth}
   \centering
     \includegraphics[width=5in]{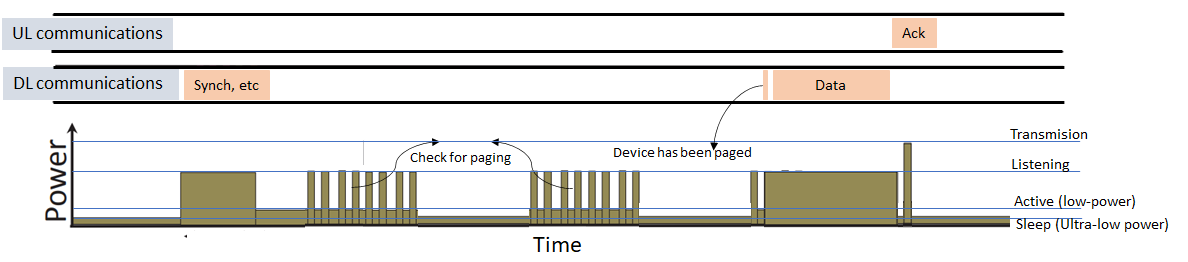}
\caption{Downlink IoT service: The device regularly wakes up and checks the DL channel for potential paging. }
\label{eng3}
    \end{subfigure}%
     \par\bigskip
    \begin{subfigure}[t]{1\textwidth}
   \centering
     \includegraphics[width=5in]{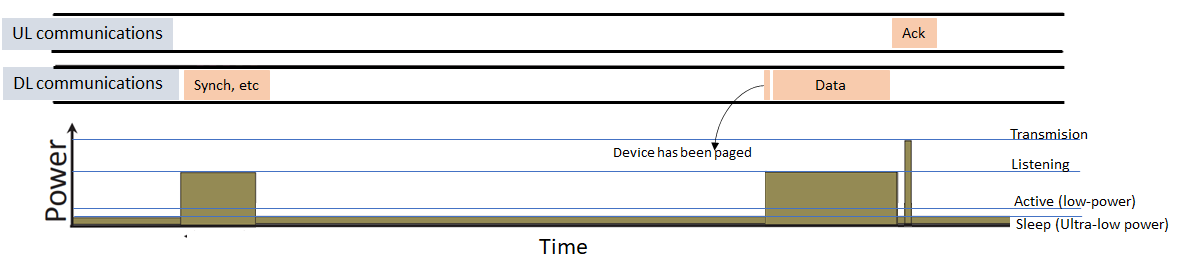}
\caption{WuR-enabled downlink IoT service: The device wakes up (by receiving the wake-up signal from the base station), receives the data, and sends an acknowledgment (ACK).}\label{eng4}
    \end{subfigure}%
    \par\bigskip
    \begin{subfigure}[t]{1\textwidth}
   \centering
     \includegraphics[width=5in]{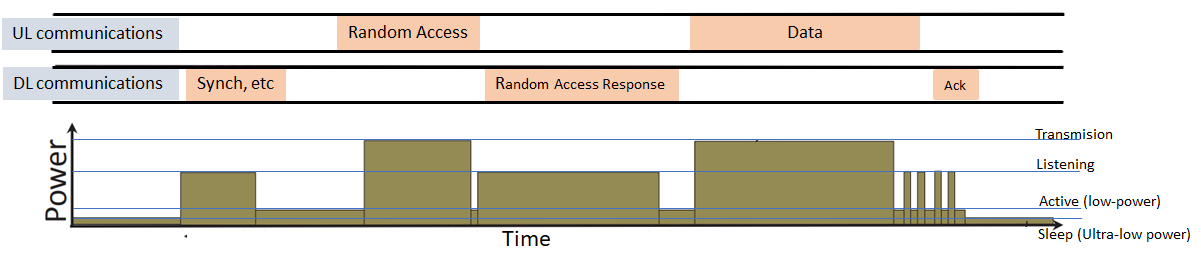}
\caption{Uplink IoT service: Upon having data to transmit, the device turns the radio on, performs synchronization, finds the random access resources, performs random access, sends data over-scheduled resources, and receives the ACK. }\label{eng1}
    \end{subfigure}%
    \par\bigskip
    \begin{subfigure}[t]{1\textwidth}
   \centering
     \includegraphics[width=5in]{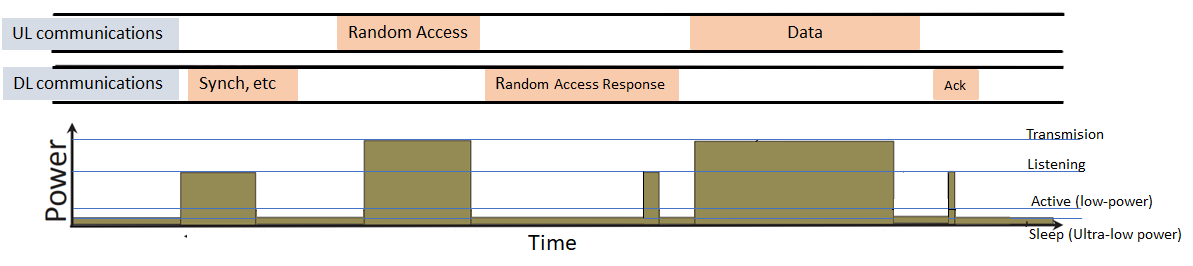}
\caption{WuR-enabled uplink IoT service: Upon having data to transmit, the device turns the radio on, performs synchronization, finds the random access resources, sends RA requests, and sleeps until the reception of the wake-up signal, which is immediately followed by data transmission and reception of the ACK.}\label{eng2}
    \end{subfigure}%
\caption{{WuR for improving the energy efficiency of uplink and downlink communications }}  \label{eng}
\end{figure*}

\section{WuRs in Cellular IoT: Opportunities, Challenges, Future Research Cirections }
In this section, we  investigate the performance of WuR-enabled cellular  IoT. To this end, we first investigate the message exchanges in up/downlink communications with/without WuRs, and compare the performance of these two schemes in terms of energy efficiency in Subsection \ref{eval}. Then, we investigate challenges in the integration of WuRs and future directions for research in Subsection \ref{rem} and Subsection \ref{futr}, respectively.  

\subsection{Integration of WuRs in Cellular IoT}\label{integ}
In many IoT applications, IoT devices are in the reporting or actuating state \cite{azari2018serving}. In the former, the device gathers data such as temperature from the environment and sends the data to the network when the network requests for the data (uplink IoT service).  In the latter, the network sends some information to the device base on which the device acts on its environment, e.g., gather data or opens a valve or so on (downlink IoT service). 

Let us investigate the downlink and uplink IoT services described above, in detail in Fig.~\ref{eng}. This figure depicts message exchanges and symbolic energy consumption levels for downlink and uplink IoT communications. First,  Fig.~\ref{eng3} depicts a simplified scenario for downlink IoT service in which, device sync itself with the network, check regularly (or in sleep/wake up cycles as in DRX) for paging, sends data as soon as is being paged by the network, and finally sends the ACK to the network. One further observes the respective power consumption levels for listening, transmission, and being in the active and sleep modes for the device. Figure \ref{eng4} depicts message exchanges and power consumption levels in the case a WuR is embedded in the device to remove the need for listening for potential paging. One observes in this figure that leveraging WuR prevents the device from regular checking of the downlink control channel at the cost of continuous operation of the WuR. In the following subsection, we will quantify the saving brought by leveraging WuRs instead of listening for paging.  

\begin{figure*}[ht]
   \centering
     \includegraphics[width=5in]{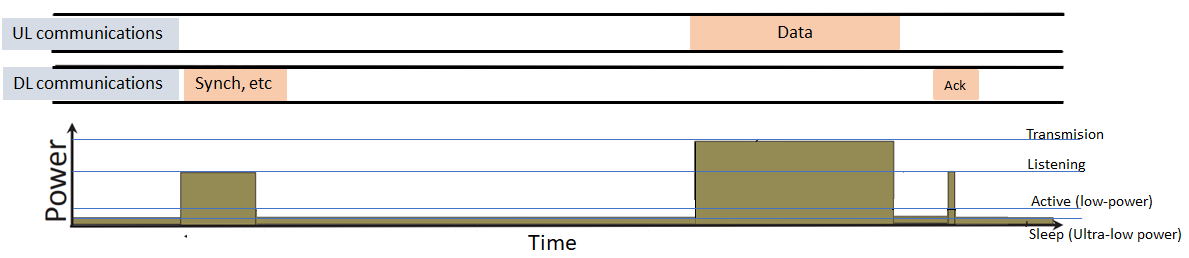}
\caption{WuR and GFA enabled IoT service: Upon having data to transmit, the device turns the radio on, performs synchronization, sends data, and sleeps until reception of the wake-up signal for the ACK or downlink message.}\label{eng5}
   \end{figure*}%

Figure \ref{eng1} presents the uplink IoT service in which, upon having data to transmit, the device turns the radio on, performs synchronization, finds the random access resources, performs random access, sends data over the scheduled resources, and receives the ACK \cite{mnbt}. One further observes the respective power consumption levels for listening, transmission, and being in the active and sleep modes for the device. In order to save energy in listening for random access and scheduling responses,  WuR-enabled uplink IoT service has been presented in Fig. \ref{eng2}. In this figure, upon having data to transmit, the device turns the radio on, performs synchronization, finds the random access resources, sends RA request, and sleeps until the reception of the wake-up signal. When the BS intends to schedule the device for data transmission (if needed), the BS sends the WuR message to the device, which is immediately followed by data transmission and reception of the ACK. In the following subsection, we  quantify the savings by leveraging WuRs instead of listening for scheduling responses.

\subsubsection{Evaluation of Energy Saving}\label{eval}

\begin{table}
\centering
	\caption{Simulation setup and parameters}
		\begin{tabular}{p{6 cm}p{2 cm}}\\
			\toprule[0.5mm]
      Parameters  &Value  \\ \hline\hline
        {Uplink transmission rate} &  $1$ kbps \\
        {Downlink transmission rate} &  $10$ kbps \\
        {Transmission power consumption} &  $80$ mW \\
        {Listening power consumption} &  $60$ mW \\
        {Active power consumption} &  $40$ mW \\
        {Sleep power consumption} &  $20$ $\mu$W \\
        {Random access duration} &  $10$ msec \\
        {Random access response duration} &  $20$ msec \\
        {Wake-up duration to check for paging signal} &  $10$ msec \\
        {Synchronization duration} &  $5$ msec \\
        {ACK duration} &  $1$ msec \\
        	\bottomrule[0.5mm]
	\end{tabular}
	\label{tab:sim}
\end{table}

Now, let us evaluate the energy-saving performance of implementing WuRs in uplink and downlink communications. We assume that the IoT device follows the downlink and uplink communication procedures according to Fig. \ref{eng4} and Fig. \ref{eng2}, respectively, and simulation parameters are summarized in Table \ref{tab:sim}. For WuR-enabled IoT devices in uplink and downlink, the device is in sleep mode unless it is paged or has a message to transmit or receive. {The wake-up duration in Table \mbox{\ref{tab:sim}}, is defined as the time in which a device listens to the communications signals after receiving a paging signal. This is similar to the the wake-up duration in DRX schemes; However, in the former the duration between the listening periods is not fixed.  The higher on-time is, the higher energy is consumed. Hence, this parameter should be adjusted considering unreachability/power consumption tradeoff as well as other restrictions of the device application. For instance,  in the application of data gathering services, the unreachability of systems with long sleep times (short wake-up duration) is acceptable because most communications are uplink-oriented. In contrast, in other applications, unreachability may cause significant costs to the system. 10 msec is a typical value for this parameter in the state-of-the-art devices and is chosen for this paper as well.}


In Fig. \ref{fig:pow_comp}, we plot the total power consumption of the IoT device in the downlink where the device is receiving data. The device wakes up according to the wake-up procedures defined in Fig. \ref{eng}.  The total power consumption of the IoT device with WuR is much lower than the IoT devices without WuR. However, as the paging rate  increases, the gaps between the power consumption of devices with and without WuR decrease. This is because with the larger arrival rate, there is not much room to sleep, and hence less energy is saved. Therefore, the impact of WuR will be smaller on the total power saving. 

In Fig. \ref{fig:DL}, we illustrate the uplink power consumption of the IoT device. We assume that the generated packets are transmitted with the uplink transmission success rate of $P_s$. As the packet generation rate increases, the uplink power consumption and the power consumption gap between the device with and without WuR increases because the IoT transmits more power to send more packets. Moreover, devices without WuR should spend more time listening to the channel for the random access response and ACK signal at each transmission. Hence more power is consumed compared to IoT device with WuR. 

In Fig. \ref{fig:tot}, we investigate the total power consumption of the IoT device in both uplink and downlink. We assumed that devices are paged every $10 s$ and $50 s$ in the downlink. The total power consumption increases when the downlink load is higher, i.e., the paging period is lower. However, in total power consumption, the uplink power consumption is dominant as the transmission power is higher than the power consumption of other modules. Increasing the uplink packet generation rate, the gap between the power consumption of devices with and without WuR increases, highlighting the benefits of using WuR in the uplink, downlink, and total power consumption.

\subsection{Remaining Challenges}\label{rem}

In this subsection, we investigate two significant challenges we encounter in the integration of WuRs in cellular IoT systems. 

\subsubsection{Interference of the wake-up and communications signals}
In most systems, the wake-up signal is sent on the same channels as used by the primary communications. Therefore, the support of WuRs either reduces the data throughput of the network or will suffer from the interference of the primary communications. To limit the negative impact on the data throughput, we might limit the maximum number of devices that expect a wake-up signal in each cell and limit the wake-up signal's time duration.
However, these limitations will affect the large-scale deployment of WuRs. Out of bound implementation of WuRs or guard band implementation are also two choices. However, they will increase the complexity of end devices, as they need to have separate radios for WuR and communications. Then, here there is a choice of more complexity at the network access side (by joint scheduling of wake-up signals and data packets) or on the device side (two bands for two purposes). 

\subsubsection{Reliability of communications: False alarms and missed alarms}
The ubiquitous coverage and reliability of communications are the key performance indicators that distinguish cellular-based solutions. While the introduction of WuRs to IoT serving procedures seems a promising method to maintain the preserved battery power and increase the device's lifetime, the reliability of WuRs, including (i) the risk of false alarm (making device awake when another device has been paged); and (ii) and risk of missed alarm (missing the paging from the network when data is waiting for the device or input from the device is needed), are to be minimized as much as possible. If such reliability concerns are not addressed properly, while saving energy for IoT devices is of paramount importance, in many applications, this saving could not be achieved at the cost of degrading the reliability of communications. 
A potential solution for these concerns consists in  generating (and continuously updating \cite{azari2019risk}) radio maps of the service area for the wake-up signals. Based on these maps and the location of devices, they could be categorized into classes, as in NB-IoT \cite{mnbt}. Then, the BS can send wake-up signals with appropriate power and a number of repetitions to the end device to ensure the device has received enough power to detect the wake-up signal.

\subsection{Further Energy Saving for WuR-enabled Cellular IoT }\label{futr}
In this subsection, we present directions for future research for the integration of WuRs into cellular IoT, including leveraging passive WuRs and integration of WuRs with grant-free radio access. 
\subsubsection{Passive WuR: Further energy saving}
A passive WuR is designed based on passive circuits which harvest energy from the incoming radio signal in the wake-up message to switch itself on and wake up the primary circuit \cite{bello2019advances}. While passive WuR does not need to be powered by a battery, the received  energy from wake-up signals has a limited range, which is a limiting factor of the system. Based on the current technology, the communication range of the passive WuRs is about 3m \cite{bello2019advances}. This range is  much less than the active ones, typically 200m in the range  \cite{bello2019advances}. 

  \begin{figure}[!t]
 
    \begin{subfigure}[t]{0.5\textwidth}
     \resizebox{.9\columnwidth}{!}{
%
%
\begin{tikzpicture}

\begin{axis}[%
width=6.028in,
height=4.754in,
at={(1.011in,0.642in)},
scale only axis,
xmode=log,
xmin=0.0001,
xmax=0.1,
xminorticks=true,
xlabel style={font=\color{white!15!black}},
xlabel={\LARGE Paging Rate},
ymode=log,
ymin=0.001,
ymax=1,
yminorticks=true,
ylabel style={font=\color{white!15!black}},
ylabel={\LARGE Power Consumption (m W)},
axis background/.style={fill=white},
xmajorgrids,
xminorgrids,
ymajorgrids,
yminorgrids,
legend style={at={(0.13,0.679)}, anchor=south west, legend cell align=left, align=left, draw=white!15!black}
]
\addplot [color=black, line width=1.5pt, mark=triangle, mark options={solid, rotate=270, black}]
  table[row sep=crcr]{%
0.0001	0.474233520000001\\
0.00213877551020408	0.48279984535\\
0.00417755102040816	0.4911361869\\
0.00621632653061225	0.4988096398\\
0.00825510204081633	0.506609336016667\\
0.0102938775510204	0.514921065719781\\
0.0123326530612245	0.523448670916667\\
0.0143714285714286	0.5300454274\\
0.0164102040816327	0.5391774763\\
0.0184489795918367	0.548561929083333\\
0.0204877551020408	0.5557139917\\
0.0225265306122449	0.557699624732235\\
0.024565306122449	0.568906655457706\\
0.0266040816326531	0.57766256384082\\
0.0286428571428571	0.583099133871908\\
0.0306816326530612	0.592568575913588\\
0.0327204081632653	0.599494967479783\\
0.0347591836734694	0.604325578541464\\
0.0367979591836735	0.609918059626851\\
0.0388367346938776	0.616396376066571\\
0.0408755102040816	0.617686715278734\\
0.0429142857142857	0.626140915692526\\
0.0449530612244898	0.634582883567729\\
0.0469918367346939	0.628762434854241\\
0.049030612244898	0.640072935185531\\
0.051069387755102	0.654287981839561\\
0.0531081632653061	0.655903889895836\\
0.0551469387755102	0.657218461673593\\
0.0571857142857143	0.656486288361866\\
0.0592244897959184	0.668237126130571\\
0.0612632653061225	0.667246818180301\\
0.0633020408163265	0.680560682268037\\
0.0653408163265306	0.673682137920134\\
0.0673795918367347	0.679573865966666\\
0.0694183673469388	0.685675688401058\\
0.0714571428571429	0.686964760909125\\
0.0734959183673469	0.687959824889843\\
0.075534693877551	0.688664176119623\\
0.0775734693877551	0.689682922456802\\
0.0796122448979592	0.690431360435614\\
0.0816510204081633	0.69734279967757\\
0.0836897959183674	0.70724909540658\\
0.0857285714285714	0.72000670209737\\
0.0877673469387755	0.720992884510189\\
0.0898061224489796	0.729223189108424\\
0.0918448979591837	0.738531852082918\\
0.0938836734693878	0.7354311921165422\\
0.0959224489795918	0.748815846104862\\
0.0979612244897959	0.753144957471054\\
0.1	0.7867329058835\\
};
\addlegendentry{\LARGE No WUR}

\addplot [color=red, dashed, line width=1.5pt, mark=o, mark options={solid, red}]
  table[row sep=crcr]{%
0.0001	0.00454661226666666\\
0.00213877551020408	0.0124383268666667\\
0.00417755102040816	0.0204666945333333\\
0.00621632653061225	0.0292466540666667\\
0.00825510204081633	0.0399055932666667\\
0.0102938775510204	0.0470115527333333\\
0.0123326530612245	0.0565772674\\
0.0143714285714286	0.0641956758666667\\
0.0164102040816327	0.0730781252\\
0.0184489795918367	0.0766652682\\
0.0204877551020408	0.0880159694019215\\
0.0225265306122449	0.0979831466\\
0.024565306122449	0.104269187666667\\
0.0266040816326531	0.110862698133333\\
0.0286428571428571	0.1204284128\\
0.0306816326530612	0.129550005\\
0.0327204081632653	0.135870209333333\\
0.0347591836734694	0.146151349901129\\
0.0367979591836735	0.155958210133333\\
0.0388367346938776	0.1650624679582\\
0.0408755102040816	0.170668071455824\\
0.0429142857142857	0.1790525784\\
0.0449530612244898	0.188037517533333\\
0.0469918367346939	0.192784206168075\\
0.049030612244898	0.205504106479788\\
0.051069387755102	0.213045028733333\\
0.0531081632653061	0.220492620866667\\
0.0551469387755102	0.229443396733333\\
0.0571857142857143	0.238346338601388\\
0.0592244897959184	0.246969152533333\\
0.0612632653061225	0.2536651528\\
0.0633020408163265	0.260566132666667\\
0.0653408163265306	0.2692436024\\
0.0673795918367347	0.272548100911441\\
0.0694183673469388	0.282100815800301\\
0.0714571428571429	0.294524419733333\\
0.0734959183673469	0.298487358666667\\
0.075534693877551	0.312618276303455\\
0.0775734693877551	0.318293379868282\\
0.0796122448979592	0.328865994038312\\
0.0816510204081633	0.334027304986702\\
0.0836897959183674	0.347535468614219\\
0.0857285714285714	0.347906931743793\\
0.0877673469387755	0.356341924679305\\
0.0898061224489796	0.373407402466666\\
0.0918448979591837	0.37773612964672\\
0.0938836734693878	0.380493346189998\\
0.0959224489795918	0.392398503901704\\
0.0979612244897959	0.4085272406\\
0.1	0.408023613197776\\
};
\addlegendentry{\LARGE $\text{WUR-2 }\mu\text{ W}$}

\addplot [color=blue, dashed, line width=1.5pt, mark=square, mark options={solid, blue}]
  table[row sep=crcr]{%
0.0001	0.002407483935\\
0.00213877551020408	0.01132451805\\
0.00417755102040816	0.019626584295\\
0.00621632653061225	0.027416177315\\
0.00825510204081633	0.0378364662233333\\
0.0102938775510204	0.044020309805\\
0.0123326530612245	0.0526640248666667\\
0.0143714285714286	0.0617860482716667\\
0.0164102040816327	0.0676965728\\
0.0184489795918367	0.075998639045\\
0.0204877551020408	0.085633135675\\
0.0225265306122449	0.0946094976567964\\
0.024565306122449	0.104423820591667\\
0.0266040816326531	0.113955822576667\\
0.0286428571428571	0.118978056850599\\
0.0306816326530612	0.125401057935\\
0.0327204081632653	0.136197160541667\\
0.0347591836734694	0.148393685454831\\
0.0367979591836735	0.153313766256667\\
0.0388367346938776	0.165100650431667\\
0.0408755102040816	0.166364751053333\\
0.0429142857142857	0.174393498245\\
0.0449530612244898	0.184610958867804\\
0.0469918367346939	0.195322232268237\\
0.049030612244898	0.201860553032223\\
0.051069387755102	0.211549884600728\\
0.0531081632653061	0.214063421350651\\
0.0551469387755102	0.227533716436963\\
0.0571857142857143	0.232473797078333\\
0.0592244897959184	0.240536709151667\\
0.0612632653061225	0.252344750268246\\
0.0633020408163265	0.259436048899921\\
0.0653408163265306	0.261035638151667\\
0.0673795918367347	0.277157123139081\\
0.0694183673469388	0.288811686946667\\
0.0714571428571429	0.293355616208333\\
0.0734959183673469	0.302033496151667\\
0.075534693877551	0.313209230128347\\
0.0775734693877551	0.315494459528333\\
0.0796122448979592	0.325578589247636\\
0.0816510204081633	0.338376414920031\\
0.0836897959183674	0.339614865985\\
0.0857285714285714	0.361680372812883\\
0.0877673469387755	0.35714145028\\
0.0898061224489796	0.37456553993\\
0.0918448979591837	0.374529033302134\\
0.0938836734693878	0.389487250862601\\
0.0959224489795918	0.397535831928348\\
0.0979612244897959	0.402226567444411\\
0.1	0.40838877278\\
};
\addlegendentry{\LARGE $\text{WUR-0.1 }\mu\text{ W}$}

\end{axis}

\end{tikzpicture}%
}
    \caption{ Downlink power consumption of IoT devices with and without WuR.  }
    \label{fig:pow_comp}
    \end{subfigure}%
    \par\bigskip
    \begin{subfigure}[t]{0.5\textwidth}
     \resizebox{.9\columnwidth}{!}{
%
%
\begin{tikzpicture}

\begin{axis}[%
width=6.028in,
height=4.754in,
at={(1.011in,0.642in)},
scale only axis,
xmin=0,
xmax=0.1,
xlabel style={font=\color{white!15!black}},
xlabel={ \LARGE Packet Generation Rate},
ymin=0,
ymax=2,
xticklabel style = {
           /pgf/number format/fixed,
        },
ylabel style={font=\color{white!15!black}},
ylabel={ \LARGE Power Consumption (m W)},
axis background/.style={fill=white},
 xmode=log,
xmajorgrids,
xminorgrids,
ymajorgrids,
yminorgrids,
legend style={at={(.60833,.929)},legend cell align=left, align=left, draw=white!15!black}
]
\addplot [color=black, line width=1.5pt, mark=triangle, mark options={solid, rotate=270, black}]
  table[row sep=crcr]{%
0.001	0.024361884168320859\\
0.00441379310344828	0.07481989787103701\\
0.00782758620689655	0.144198437116218\\
0.0112413793103448	0.176280099677282\\
0.0146551724137931	0.246099693026121\\
0.0180689655172414	0.313496698298203\\
0.0214827586206897	0.365202485963436\\
0.0248965517241379	0.44303812524273\\
0.0283103448275862	0.502213196021951\\
0.0317241379310345	0.559270429098736\\
0.0351379310344828	0.624062815629291\\
0.038551724137931	0.677757135145595\\
0.0419655172413793	0.766823806068586\\
0.0453793103448276	0.802484726810136\\
0.0487931034482759	0.87402050905407\\
0.0522068965517241	0.918997654717594\\
0.0556206896551724	0.962307047511457\\
0.0590344827586207	1.04541611729432\\
0.062448275862069	1.10064399606163\\
0.0658620689655172	1.17091280396406\\
0.0692758620689655	1.22851791344781\\
0.0726896551724138	1.2856874921212\\
0.0761034482758621	1.35284033737594\\
0.0795172413793103	1.42360101364639\\
0.0829310344827586	1.4757944260552\\
0.0863448275862069	1.55173812854514\\
0.0897586206896552	1.59584161804746\\
0.0931724137931035	1.66004955591749\\
0.0965862068965517	1.74037587174975\\
0.1	1.77910655580283\\
};
\addlegendentry{\LARGE $\text{No WUR- P}_\text{s}\text{: 0.9}$}

\addplot [color=black, line width=1.5pt, mark=square, mark options={solid, black}]
  table[row sep=crcr]{%
0.001	0.02727247841551281\\
0.00441379310344828	0.0820494549049706\\
0.00782758620689655	0.158658443969765\\
0.0112413793103448	0.226175352740677\\
0.0146551724137931	0.290181718947735\\
0.0180689655172414	0.341612342696143\\
0.0214827586206897	0.39994827300062286\\
0.0248965517241379	0.475050045850502\\
0.0283103448275862	0.54032266112775\\
0.0317241379310345	0.601144801634074\\
0.0351379310344828	0.682215478299153\\
0.038551724137931	0.739872215586452\\
0.0419655172413793	0.799170302808169\\
0.0453793103448276	0.874313817318249\\
0.0487931034482759	0.925615360867585\\
0.0522068965517241	1.00224241220583\\
0.0556206896551724	1.05297296695007\\
0.0590344827586207	1.1142628823768\\
0.062448275862069	1.18548638433989\\
0.0658620689655172	1.25304853808616\\
0.0692758620689655	1.29581195579802\\
0.0726896551724138	1.39055752426421\\
0.0761034482758621	1.47005166308296\\
0.0795172413793103	1.48473736674147\\
0.0829310344827586	1.54810592904511\\
0.0863448275862069	1.63568993433201\\
0.0897586206896552	1.6954510894317\\
0.0931724137931035	1.75353124898825\\
0.0965862068965517	1.8405214500339\\
0.1	1.90662283912142\\
};
\addlegendentry{\LARGE $\text{No WUR- P}_\text{s}\text{: 0.75}$}

\addplot [color=red, dashdotted, line width=1.5pt, mark=triangle, mark options={solid, rotate=270, red}]
  table[row sep=crcr]{%
0.001	0.0281867323824488\\
0.00441379310344828	0.0720563526810919\\
0.00782758620689655	0.111673393758968\\
0.0112413793103448	0.134369694396682\\
0.0146551724137931	0.183763845277483\\
0.0180689655172414	0.231444126523045\\
0.0214827586206897	0.275098133084311\\
0.0248965517241379	0.330163267755838\\
0.0283103448275862	0.357877828440358\\
0.0317241379310345	0.398243196444649\\
0.0351379310344828	0.444080829732305\\
0.038551724137931	0.482067091072717\\
0.0419655172413793	0.545077664050997\\
0.0453793103448276	0.570306125220282\\
0.0487931034482759	0.620914400806748\\
0.0522068965517241	0.652733662600383\\
0.0556206896551724	0.683373065948491\\
0.0590344827586207	0.742168910145228\\
0.062448275862069	0.781240094118323\\
0.0658620689655172	0.830952043633773\\
0.0692758620689655	0.87170500899593\\
0.0726896551724138	0.912149856342381\\
0.0761034482758621	0.959657405606034\\
0.0795172413793103	1.00971732936794\\
0.0829310344827586	1.04664176796534\\
0.0863448275862069	1.10036844437969\\
0.0897586206896552	1.13156963462386\\
0.0931724137931035	1.17699379736642\\
0.0965862068965517	1.23382097113831\\
0.1	1.26122114865405\\
};
\addlegendentry{\LARGE $\text{WUR power: 2 }\mu\text{ W - P}_\text{s}\text{: 0.9}$}

\addplot [color=blue, dashed, line width=1.5pt, mark=triangle, mark options={solid, rotate=270, blue}]
  table[row sep=crcr]{%
0.001	0.0262876603906938\\
0.00441379310344828	0.0701589639001442\\
0.00782758620689655	0.109777525023812\\
0.0112413793103448	0.132474696484165\\
0.0146551724137931	0.181870742543634\\
0.0180689655172414	0.229552853209285\\
0.0214827586206897	0.273208534708587\\
0.0248965517241379	0.328275782145831\\
0.0283103448275862	0.355991406196018\\
0.0317241379310345	0.396358322958269\\
0.0351379310344828	0.442197714966407\\
0.038551724137931	0.480185433782065\\
0.0419655172413793	0.543198424380481\\
0.0453793103448276	0.568427853527565\\
0.0487931034482759	0.619038070876823\\
0.0522068965517241	0.650858553527283\\
0.0556206896551724	0.681499132462836\\
0.0590344827586207	0.740297232566957\\
0.062448275862069	0.779369915642134\\
0.0658620689655172	0.829083772529704\\
0.0692758620689655	0.869838301521345\\
0.0726896551724138	0.910284700675258\\
0.0761034482758621	0.957794072731537\\
0.0795172413793103	1.0078559172168\\
0.0829310344827586	1.04478177254892\\
0.0863448275862069	1.09851051037437\\
0.0897586206896552	1.1297128977609\\
0.0931724137931035	1.17513880335971\\
0.0965862068965517	1.23196815750409\\
0.1	1.25936938632308\\
};
\addlegendentry{\LARGE $\text{WUR power: 2 }\mu\text{ W - P}_\text{s}\text{: 0.75}$}

\addplot [color=red, dashdotted, line width=1.5pt, mark=square, mark options={solid, red}]
  table[row sep=crcr]{%
0.001	0.02537448955748249\\
0.00441379310344828	0.060303484310891\\
0.00782758620689655	0.114187086878871\\
0.0112413793103448	0.161675695242485\\
0.0146551724137931	0.206695132786618\\
0.0180689655172414	0.242869311870729\\
0.0214827586206897	0.280298517612682\\
0.0248965517241379	0.329690300110741\\
0.0283103448275862	0.382633952485807\\
0.0317241379310345	0.425413737030473\\
0.0351379310344828	0.482435506599102\\
0.038551724137931	0.522988877101454\\
0.0419655172413793	0.564696705437291\\
0.0453793103448276	0.617549553658508\\
0.0487931034482759	0.653632943055519\\
0.0522068965517241	0.707529249881523\\
0.0556206896551724	0.743211029307952\\
0.0590344827586207	0.78631982743458\\
0.062448275862069	0.836415500541987\\
0.0658620689655172	0.88393593235478\\
0.0692758620689655	0.914013956259851\\
0.0726896551724138	0.980654080592937\\
0.0761034482758621	1.03656697847716\\
0.0795172413793103	1.04689629660322\\
0.0829310344827586	1.09146712919293\\
0.0863448275862069	1.15307010542068\\
0.0897586206896552	1.19510363659901\\
0.0931724137931035	1.23595482485818\\
0.0965862068965517	1.29714014367876\\
0.1	1.34363313389132\\
};
\addlegendentry{\LARGE $\text{WUR power: 0.1 }\mu\text{ W - P}_\text{s}\text{: 0.9}$}

\addplot [color=blue, dashed, line width=1.5pt, mark=square, mark options={solid, blue}]
  table[row sep=crcr]{%
0.001	0.02518468885354094\\
0.00441379310344828	0.0584057400322812\\
0.00782758620689655	0.112291501372429\\
0.0112413793103448	0.159782012301595\\
0.0146551724137931	0.204803253487358\\
0.0180689655172414	0.240978881840027\\
0.0214827586206897	0.27840958713145\\
0.0248965517241379	0.327803348443105\\
0.0283103448275862	0.380749121932532\\
0.0317241379310345	0.423530620390222\\
0.0351379310344828	0.480554674457349\\
0.038551724137931	0.521109669674519\\
0.0419655172413793	0.562819168976934\\
0.0453793103448276	0.615674134674569\\
0.0487931034482759	0.651758969702774\\
0.0522068965517241	0.705657435809924\\
0.0556206896551724	0.7413406447776\\
0.0590344827586207	0.784451169998726\\
0.062448275862069	0.834548850120136\\
0.0658620689655172	0.882071185773443\\
0.0692758620689655	0.912150414713029\\
0.0726896551724138	0.97879320889072\\
0.0761034482758621	1.03470834684802\\
0.0795172413793103	1.04503807880396\\
0.0829310344827586	1.08961069706256\\
0.0863448275862069	1.15121614132854\\
0.0897586206896552	1.19325135652228\\
0.0931724137931035	1.23410418142792\\
0.0965862068965517	1.29529195155386\\
0.1	1.3417868044439\\
};
\addlegendentry{\LARGE $\text{WUR power: 0.1 }\mu\text{ W - P}_\text{s}\text{: 0.9}$}

\end{axis}

\end{tikzpicture}%
}
    \caption{ Uplink power consumption of IoT devices with and without WuR.}
    \label{fig:DL}
    \end{subfigure}%
    \par\bigskip
    \begin{subfigure}[t]{0.5\textwidth}
\resizebox{.9\columnwidth}{!}{
%
%
\definecolor{mycolor1}{rgb}{0.00000,0.44700,0.74100}%
\definecolor{mycolor2}{rgb}{0.85000,0.32500,0.09800}%
\definecolor{mycolor3}{rgb}{0.92900,0.69400,0.12500}%
\definecolor{mycolor4}{rgb}{0.49400,0.18400,0.55600}%
\begin{tikzpicture}

\begin{axis}[%
width=6.028in,
height=4.754in,
at={(1.011in,0.642in)}, 
scale only axis,
xmode=log,
ymode=log,
xmin=0.001,
xmax=0.1,
xminorticks=true,
xlabel style={font=\color{white!15!black}},
xlabel={\LARGE  Uplink Packet Generation Rate},
ymin=0,
ymax=3,
ylabel style={font=\color{white!15!black}},
ylabel={\LARGE  Uplink + Downlink Power Consumption (m W)},
axis background/.style={fill=white},
xmajorgrids,
xminorgrids,
ymajorgrids,
yminorgrids,
legend style={at={(0.71,0.9)},legend cell align=left, align=left, draw=white!15!black}
]
\addplot [color=mycolor1, line width=1.5pt, mark=o, mark options={solid, mycolor1}]
  table[row sep=crcr]{%
0.001	0.789518534848895\\
0.00441379310344828	0.846654018306457\\
0.00782758620689655	0.893714565527738\\
0.0112413793103448	0.961977742144245\\
0.0146551724137931	1.01896075549823\\
0.0180689655172414	1.08511419342833\\
0.0214827586206897	1.13868114362629\\
0.0248965517241379	1.20208798415038\\
0.0283103448275862	1.25766260051494\\
0.0317241379310345	1.31622180255511\\
0.0351379310344828	1.38528502237647\\
0.038551724137931	1.45543779040504\\
0.0419655172413793	1.48910862461202\\
0.0453793103448276	1.57693301178261\\
0.0487931034482759	1.6006906863164\\
0.0522068965517241	1.67174549937675\\
0.0556206896551724	1.74172757810961\\
0.0590344827586207	1.82691589005732\\
0.062448275862069	1.86483340212429\\
0.0658620689655172	1.92319853044398\\
0.0692758620689655	1.99146053106873\\
0.0726896551724138	2.06252942411751\\
0.0761034482758621	2.11228902769647\\
0.0795172413793103	2.16860472231119\\
0.0829310344827586	2.21004951713322\\
0.0863448275862069	2.2544782109783\\
0.0897586206896552	2.35378248434653\\
0.0931724137931035	2.4098950769135\\
0.0965862068965517	2.44027153626012\\
0.1	2.55031744699845\\
};
\addlegendentry{\LARGE  No WuR- Downlink paging every 10s}

\addplot [color=mycolor2, dashdotted, line width=1.5pt, mark=square, mark options={solid, mycolor2}]
  table[row sep=crcr]{%
0.001	0.78703239259095\\
0.00441379310344828	0.827453119140445\\
0.00782758620689655	0.860746291708309\\
0.0112413793103448	0.90903934933974\\
0.0146551724137931	0.949352210303948\\
0.0180689655172414	0.996152724854033\\
0.0214827586206897	1.03404887826018\\
0.0248965517241379	1.07890630147767\\
0.0283103448275862	1.11822278635232\\
0.0317241379310345	1.15965072831382\\
0.0351379310344828	1.20850977971399\\
0.038551724137931	1.25813963635604\\
0.0419655172413793	1.28196020283789\\
0.0453793103448276	1.34409191720129\\
0.0487931034482759	1.36089937917884\\
0.0522068965517241	1.41116739119052\\
0.0556206896551724	1.46067649300686\\
0.0590344827586207	1.52094330539865\\
0.062448275862069	1.54776820113678\\
0.0658620689655172	1.58905884486763\\
0.0692758620689655	1.63735107053898\\
0.0726896551724138	1.68762904349485\\
0.0761034482758621	1.72283167429031\\
0.0795172413793103	1.7626724379404\\
0.0829310344827586	1.79199272398726\\
0.0863448275862069	1.82342398134587\\
0.0897586206896552	1.89367718716664\\
0.0931724137931035	1.93337426545981\\
0.0965862068965517	1.95486421325562\\
0.1	2.0327166334447\\
};
\addlegendentry{\LARGE  WuR-Downlink paging every 10s}

\addplot [color=mycolor3, line width=1.5pt, mark=o, mark options={solid, mycolor3}]
  table[row sep=crcr]{%
0.001	0.651795767203505\\
0.00441379310344828	0.708931250661067\\
0.00782758620689655	0.755991797882348\\
0.0112413793103448	0.824254974498856\\
0.0146551724137931	0.881237987852837\\
0.0180689655172414	0.947391425782943\\
0.0214827586206897	1.0009583759809\\
0.0248965517241379	1.06436521650499\\
0.0283103448275862	1.11993983286955\\
0.0317241379310345	1.17849903490972\\
0.0351379310344828	1.24756225473108\\
0.038551724137931	1.31771502275965\\
0.0419655172413793	1.35138585696663\\
0.0453793103448276	1.43921024413722\\
0.0487931034482759	1.46296791867101\\
0.0522068965517241	1.53402273173136\\
0.0556206896551724	1.60400481046422\\
0.0590344827586207	1.68919312241193\\
0.062448275862069	1.7271106344789\\
0.0658620689655172	1.78547576279859\\
0.0692758620689655	1.85373776342334\\
0.0726896551724138	1.92480665647212\\
0.0761034482758621	1.97456626005108\\
0.0795172413793103	2.0308819546658\\
0.0829310344827586	2.07232674948783\\
0.0863448275862069	2.11675544333291\\
0.0897586206896552	2.21605971670114\\
0.0931724137931035	2.27217230926811\\
0.0965862068965517	2.30254876861473\\
0.1	2.41259467935306\\
};
\addlegendentry{ \LARGE No Wur-Downlink paging every 50s}

\addplot [color=mycolor4, dashdotted, line width=1.5pt, mark=square, mark options={solid, mycolor4}]
  table[row sep=crcr]{%
0.001	0.188138439178851\\
0.00441379310344828	0.228559165728345\\
0.00782758620689655	0.261852338296209\\
0.0112413793103448	0.310145395927641\\
0.0146551724137931	0.350458256891849\\
0.0180689655172414	0.397258771441934\\
0.0214827586206897	0.435154924848079\\
0.0248965517241379	0.480012348065569\\
0.0283103448275862	0.519328832940218\\
0.0317241379310345	0.560756774901722\\
0.0351379310344828	0.609615826301894\\
0.038551724137931	0.659245682943945\\
0.0419655172413793	0.683066249425794\\
0.0453793103448276	0.745197963789186\\
0.0487931034482759	0.762005425766742\\
0.0522068965517241	0.812273437778419\\
0.0556206896551724	0.86178253959476\\
0.0590344827586207	0.922049351986552\\
0.062448275862069	0.948874247724676\\
0.0658620689655172	0.990164891455529\\
0.0692758620689655	1.03845711712688\\
0.0726896551724138	1.08873509008275\\
0.0761034482758621	1.12393772087821\\
0.0795172413793103	1.1637784845283\\
0.0829310344827586	1.19309877057516\\
0.0863448275862069	1.22453002793377\\
0.0897586206896552	1.29478323375454\\
0.0931724137931035	1.33448031204771\\
0.0965862068965517	1.35597025984352\\
0.1	1.4338226800326\\
};
\addlegendentry{\LARGE  WuR-Downlink paging every 50s}

\end{axis}
\end{tikzpicture}%
}
    \caption{ Total power consumption of IoT devices with and without WuR.  }
    \label{fig:tot}
    \end{subfigure}%
\caption{{WuR for improving energy efficiency of uplink and downlink communications }}  \label{perfM}
\end{figure}
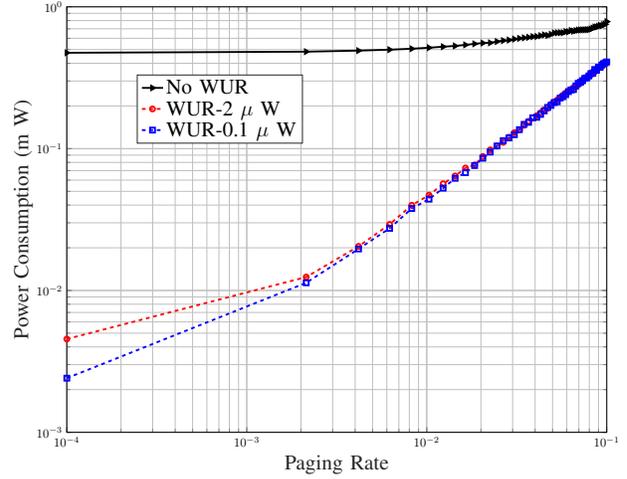
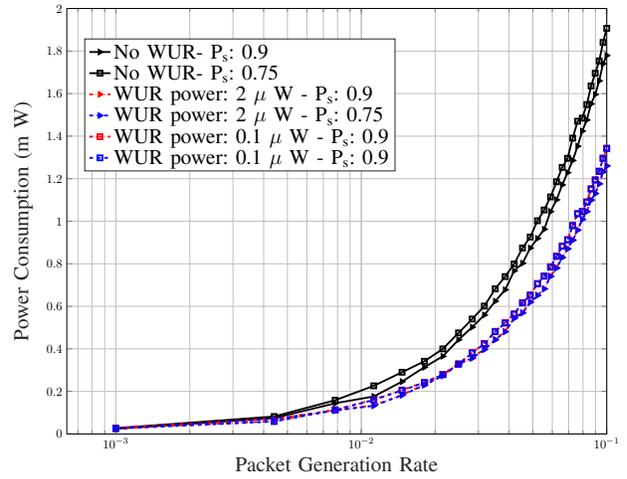
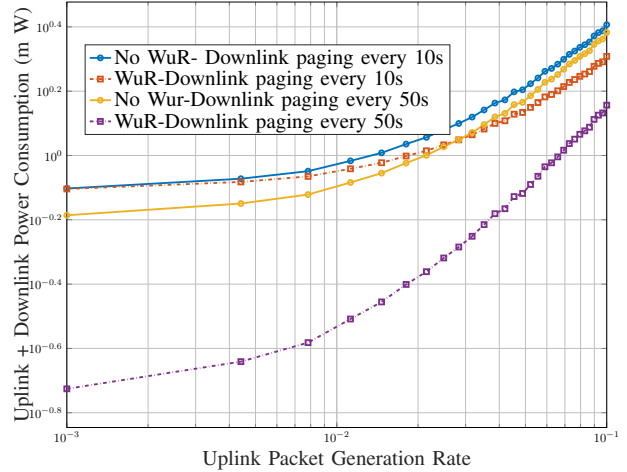

\subsubsection{WuR meets GFA for a close-to-zero energy profile in the idle mode}
The works in \cite{mads,nbt} have shown that energy consumption in the signaling states and resource reservation can be as high as in the  transmission state (due to the period of being in the listening mode).
A potential solution to decrease the overall energy consumption in resource reservation is to use grant-free access (GFA), especially in the uplink communications where WuR cannot save much energy (Fig. \ref{eng2} and \ref{fig:DL}) ~\cite{znbt,gf}. 
The GFA scheme has already been implemented in IoT technologies that operate in the unlicensed spectrum, such as SigFox and  LoRaWAN~\cite{mag_all}. In GFA, once a packet arrives at the device, it is transmitted without any handshaking and resource reservation. As part of the efforts shaping the  5G new radio (NR) and beyond, using GFA has become a hot topic in recent years {\mbox{~\cite{s20gr}}}. Fig. \ref{eng5} represents a scenario in which the IoT device leverages both GFA and WuR  for uplink and downlink communications. As one observes in this figure, upon having data to transmit, the device turns the radio on, performs synchronization, sends data, and sleeps until reception of the wake-up signal for the ACK or a potential downlink message. Despite the energy benefits in GFA, there are some challenges to be dealt with before the realization of GFA, mainly in terms of resource efficiency and reliability of communications. GFA has been investigated in a several recent works \cite{zgf1,anf,dln} and  the 3GPP \cite{gf3,rsma}.
In the context of the 3GPP standardization, the set of radio resources that should be allocated to GFA, the choice of modulation and coding scheme, and the impact on grant-based communications  are the main study items \cite{gf3,rsma}.

\section{Conclusion}
In this work, we have investigated the integration of WuRs in the uplink/downlink cellular IoT communications. First, we explored the address decoding problem in the design of energy-efficient yet reliable WuRs. Then, we  designed a novel address decoder by sequential clocking of  address decoders. 
Closed-form expressions of the key performance indicators have been derived in address decoding, including reliability, delay, and power consumption in operation.  A case study, consisting of data gathering IoT devices has also been introduced to investigate the potential of the proposed WuR-based solutions. Using this system-level analysis, regions of traffic load in which the introduction of WuR in the radio of the IoT device is beneficial to have been explored. Then, device-level analysis has been carried out, where performance evaluation results show significant advantages in power efficiency. These promising results promote the integration of WuR  in future cellular networks, along with the legacy downlink control channel monitoring  schemes used in the existing  LTE and NB-IoT systems.

\ifCLASSOPTIONcaptionsoff
  \newpage
\fi

\bibliographystyle{IEEEtran}

\bibliography{refrences.bib}

\end{document}